\documentclass[prx,twocolumn,preprintnumbers,superscriptaddress,longbibliography]{revtex4-2} 

\usepackage{graphicx}
\usepackage{bm}
\usepackage{amsmath}
\usepackage{amssymb}
\usepackage{amsfonts}
\usepackage{fancyhdr}
\usepackage{latexsym,epsfig,bbm}
\usepackage{algorithm}
\usepackage{algorithmic}

\newcommand{\bra}[1]{\langle{#1}|}
\newcommand{\ket}[1]{|{#1}\rangle}

\newcommand{\braket}[2]{\langle{#1}|{#2}\rangle}

\newcommand{\figref}[1]{Fig.~\ref{#1}}

\newcommand{\Tr}{\mathrm{Tr}}
\usepackage{color}
\definecolor{blue}{rgb}{0,0.2,1}

\definecolor{red}{rgb}{0.9,0,0}

\newcommand{\past}[1]{\overleftarrow{#1}}
\newcommand{\fut}[1]{\overrightarrow{#1}}

\begin{document}

\title{Quantum adaptive agents with efficient long-term memories}

\author{Thomas J.~Elliott}
\email{physics@tjelliott.net}
\affiliation{Department of Mathematics, Imperial College London, London SW7 2AZ, United Kingdom}
\affiliation{Complexity Institute, Nanyang Technological University, Singapore 637335}
\affiliation{School of Physical and Mathematical Sciences, Nanyang Technological University, Singapore 637371}
\author{Mile Gu}
\email{mgu@quantumcomplexity.org}
\affiliation{School of Physical and Mathematical Sciences, Nanyang Technological University, Singapore 637371}
\affiliation{Complexity Institute, Nanyang Technological University, Singapore 637335}
\affiliation{Centre for Quantum Technologies, National University of Singapore, 3 Science Drive 2, Singapore 117543}
\author{Andrew J.~P.~Garner}
\affiliation{Institute for Quantum Optics and Quantum Information, Austrian Academy of Sciences, Boltzmanngasse 3, Vienna 1090, Austria}
\affiliation{School of Physical and Mathematical Sciences, Nanyang Technological University, Singapore 637371}
\author{Jayne Thompson}
\email{thompson.jayne2@gmail.com}
\affiliation{Centre for Quantum Technologies, National University of Singapore, 3 Science Drive 2, Singapore 117543}

\date{\today}

\begin{abstract}
Central to the success of adaptive systems is their ability to interpret signals from their environment and respond accordingly -- they act as agents interacting with their surroundings. Such agents typically perform better when able to execute increasingly complex strategies. This comes with a cost: the more information the agent must recall from its past experiences, the more memory it will need. Here we investigate the power of agents capable of quantum information processing. We uncover the most general form a quantum agent need adopt to maximise memory compression advantages, and provide a systematic means of encoding their memory states. We show these encodings can exhibit extremely favourable scaling advantages relative to memory-minimal classical agents, particularly when information must be retained about events increasingly far into the past.
\end{abstract}
\maketitle 

\section{Introduction}

The world is awash with complex, interacting systems. Predators chasing prey, investors trading stocks, grandmasters playing chess: all share in common that they process information from their environment and act in response, with an eye to achieving some desired outcome. They can be described as adaptive agents~\cite{bonabeau2002agent, macy2002factors, macal2005tutorial, wooldridge2009introduction, an2012modeling}, systems that receive input stimuli and respond with output actions. This framework can be applied to a plethora of problems, including financial markets~\cite{farmer2009economy, thurner2012leverage}, biofilm formation~\cite{lardon2011idynomics}, and HIV spread~\cite{mei2010complex}.

To be effective, an agent must typically adapt its future behaviour based on past experiences.  A rudimentary chatbot, for example, would base its response purely on the last phrase it heard -- often resulting in wildly out-of-context output. Meanwhile, a more sophisticated design would extract context from conversational history -- in both what they have heard, and what they have said. Tracking this contextual data requires a memory, and a policy for deciding on what action to take based on the current stimulus and this memory. For agents performing elaborate tasks, effective strategies often require copious information about past data~\cite{cassandra1994acting}; tools that ameliorate the amount of information agents must retain can thus provide a valuable competitive advantage.

\begin{figure}
\includegraphics[width=0.94\linewidth]{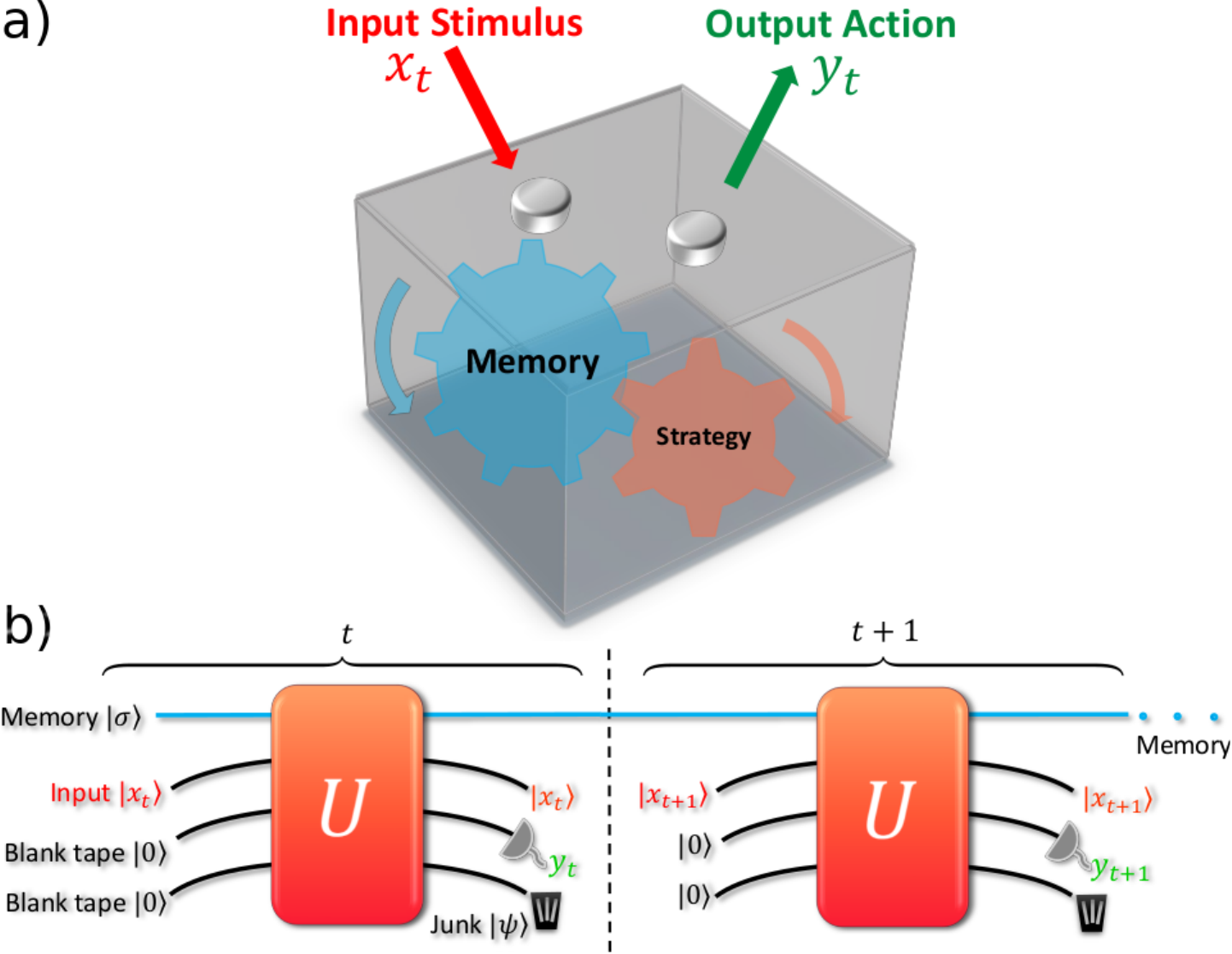}
\caption{{\bf Agents and their quantum realisations.} (a) We consider agents that alternately receive input stimuli and perform output actions. To execute complex behaviour, an agent requires a memory to keep track of relevant information about past events (both stimuli and actions), and a strategy for deciding on future actions based on this information together with the current stimulus. (b) A quantum circuit implementing a quantum agent that encompasses all memory-minimal agents (see Theorem 1). At each timestep it interacts with an input stimulus encoded in $\ket{x_t}$ , and some blank tape. After the interaction, measurement of the output tape delivers the appropriate action $y_t$. In general an agent must also dispose of additional redundant information, requiring junk tape that is discarded into the environment. This process can be repeated to execute the desired strategic behaviour ad-infinitum.}
\label{figschematic}
\end{figure}

To what extent can agents benefit from quantum technologies? Proof-of-principle quantum agents have demonstrated memory compression beyond classical bounds~\cite{thompson2017using}, yet do not make use of the full gamut of possible quantum effects. Here we identify the features of vastly improved quantum adaptive agents that use less memory -- and provide a systematic procedure for their design -- using insights from  quantum stochastic modelling~\cite{gu2012quantum, mahoney2016occam, aghamohammadi2018extreme, elliott2018superior, binder2018practical, elliott2019memory, thompson2018causal, elliott2020extreme, liu2019optimal}. The resulting agents can display extreme scaling advantages over provably minimal classical counterparts ~\cite{barnett2015computational}. We derive sufficient conditions under which such scaling advantages can occur, and illustrate this with a family of scenarios where the agent's decisions rely on events in the distant past. Complementing techniques for quantum agents to speed-up the learning of effective strategies~\cite{dong2008quantum, paparo2014quantum, dunjko2018machine}, our work illustrates that they will also be able to execute them with lower memory overhead. Together, they represent key components of quantum-enhanced artificial intelligences.

\section{Framework}

\noindent \textbf{Agents and strategies.} We describe adaptive agents as automatons that interact with their environment at discrete timesteps $t\in\mathbb{Z}$. At each timestep the agent receives an input stimulus $x_t\in\mathcal{X}$ and responds with output action $y_t\in\mathcal{Y}$, manifest by random variables $X_t$ and $Y_t$ respectively (throughout, upper case indicates random variables and lower case the corresponding variates). Taking $t=0$ as the present, we denote the past sequences of stimuli and actions as $\past{x}:=\ldots x_{-2}x_{-1}$ and $\past{y}:=\ldots y_{-2}y_{-1}$ respectively. For shorthand we denote the pair $z:=(x,y)$, and similarly $\past{z}:=(\past{x},\past{y})$ for the entire history. The agent's choice of action is governed by a strategy, describing the probability that the agent should select action $y$ in response to stimuli $x$ given preceding stimuli and actions $\past{z}$~\footnote{Here, we assume that agents must be causal and non-anticipatory, such that their decisions and internal memory can only be based on past information, and not as-yet-unseen future information. See Technical Appendix A for details.}. Each strategy $\mathcal{P}$ is thus defined by the distribution $P(Y|\past{Z},X)$; we assume strategies to be time-invariant~\cite{barnett2015computational, thompson2017using}.

To execute a desired strategy $\mathcal{P}$, an agent must be able to execute actions in a manner statistically faithful to the distribution for any sequence of received stimuli. This necessitates that the agent possesses a memory system $\mathsf{M}$ that stores relevant information from the past. A brute-force approach would be to record all past stimuli and actions, allowing a direct sampling from $P(Y|\past{z},x)$. However, storing the entire history fast becomes prohibitively expensive.

A more refined approach is to use an encoding function $f$ that maps possible histories $\{\past{z}\}$ to a corresponding memory state from the set $\{\sigma_m\}$, labelled by $m\in\mathcal{M}$. Given a history $\past{z}$, upon receiving any of the possible stimuli $x$ the agent must be able to use its memory to
\begin{enumerate}
\item Produce output $y$ with probability $P(Y|\past{z},x)$; and
\item Update the state of $\mathsf{M}$ to one consistent with the new history $\past{z}z$ (i.e, $f(\past{z}z)$).
\end{enumerate}
This process is illustrated schematically in \figref{figschematic}(a). This requires the agent to have a policy $\Lambda$ -- a systematic procedure that governs the internal dynamics of the agent. Repeated application of $\Lambda$ then allows the agent to execute the strategy over multiple timesteps. Provided such a $\Lambda$ exists for an encoding function $f$, then this can be used to specify an adaptive agent. That is, the tuple $(\mathcal{X},\mathcal{Y},\{\sigma_m\},f,\Lambda)$ formally defines an adaptive agent; see Technical Appendix A for further details.

Since the encoding function is a deterministic mapping from histories to memory states, we are able to succintly describe the update of the memory according to an update rule $m'=\lambda(z,m)$, where $\sigma_m$ is the memory state corresponding to any given history $\past{z}$, and $\sigma_{m'}$ that of $\past{z}z$. We can also replace the distribution $P(Y|\past{Z},X)$ by $P(Y|M,X)$, where the substitution of histories by memory state labels is done in accordance with the encoding function (i.e., $f(\past{z})=\sigma_m$ implies the substitution $\past{z}\to m$).

\noindent \textbf{Memory costs.} Different choices of $f$ lead to different memory states, and consequently, agents with different memory requirements. Here we are concerned with memory-minimal agents -- those that are able to extract and store the minimal amount of historical information possible whilst still being able to execute a given strategy for any future stimuli. Correspondingly, we take the amount of information stored in the agent's memory system $\mathsf{M}$ as our metric of performance:
\begin{equation}
\label{eqstatmem}
C_{f,\mathcal{R}}:=S_{\textrm{vN}}[\sigma_M],
\end{equation}
where $S_{\textrm{vN}}$ is the von Neumann entropy~\cite{nielsen2000quantum} (reducing to the Shannon entropy for classical memory states) of the memory state distribution, here assumed to be their steady-state distribution~\cite{barnett2015computational, thompson2017using}. The second subscript $\mathcal{R}$ recognises that this distribution typically depends on how the stimuli the agent receives are selected -- be they drawn from a stochastic process, or more generally, by another agent responding to the actions of the agent. The procedure for how the input stimuli are selected is referred to as the input strategy $\mathcal{R}$, as is formally defined in Technical Appendix A. It is often useful to also consider a `worst-case' information cost -- the necessary amount of memory an agent must have available to be able to respond appropriately to any input strategy.

\noindent \textbf{Memory-minimal classical agents.} Using tools from complexity science~\cite{crutchfield1989inferring, shalizi2001computational, crutchfield2012between}, the provably memory-minimal classical adaptive agents can systematically be determined~\cite{barnett2015computational}. Consider that if the strategy dictates that two histories $\past{z}$ and $\past{z}'$ should have statistically identical action responses for all possible future stimuli sequences, there should be no need to distinguish between them in the memory. Similarly, consider that if they do have different action responses, then they have to be represented by different memory states. This rationale, while seemingly simple, directly motivates an encoding function that can be shown to be memory-minimal in the design of classical adaptive agents.

This encoding function $f_\varepsilon$ is thus defined by:
\begin{equation}
\label{eqcausalencoding}
f_\varepsilon(\past{z})=f_\varepsilon(\past{z}')\Leftrightarrow P(\fut{Y}|\past{z},\fut{x})=P(\fut{Y}|\past{z}',\fut{x})\forall\fut{x}.
\end{equation}
The corresponding memory states $\{\sigma_s\}$, labelled by $s\in\mathcal{S}$ -- referred to as the causal states of the strategy~\cite{crutchfield1989inferring, barnett2015computational} -- are a partitioning of histories into equivalence classes based on their responses to future stimuli. The respective information cost Eq.~\eqref{eqstatmem} of this encoding function is given by \mbox{$C_{\mu,\mathcal{R}}=-\sum_{s\in\mathcal{S}}P(s)\log_2[P(s)]$}, where \mbox{$P(s)=\sum_{\past{z}\in s}P(\past{z})$} for the given input strategy $\mathcal{R}$.

The agent as a whole is called the $\varepsilon$-transducer of the strategy~\cite{barnett2015computational}, and crucially, is classically memory-minimal for any non-pathological input strategy. In recognition of this, its memory requirements are seen as fundamental properties of the strategy; in particular, the worst-case information cost is designated as the structural complexity of the strategy~\cite{barnett2015computational}. These ideas have seen application in contexts such as agent-based learning~\cite{marzen2018optimized, zhang2019learning} and energy-harvesting~\cite{boyd2017leveraging}, and understanding quantum contextuality~\cite{cabello2018optimal}.

\section{Quantum adaptive agents}

A quantum adaptive agent is able to store and process quantum information in its memory system $\mathsf{M}$, such that the encoding function $f$ maps histories into quantum states $\{\rho_m\}$, and the policy $\Lambda$ is a quantum channel. As per Eq.~\eqref{eqstatmem}, the information cost of a quantum encoding function $q$ is given by $C_{q,\mathcal{R}}=-\mathrm{Tr}({\rho\log_2[\rho]})$, where $\rho=\sum_mP(m)\rho_m$. A specific design for a quantum agent has already demonstrated the potential for a quantum memory advantage over memory-minimal classical agents~\cite{thompson2017using}. Yet, there is a great flexibility in how a quantum agent can be designed beyond these prior proof-of-principle constructions; we now proceed to explore how quantum agents can maximise their advantage.

A central result of this work (proven in Technical Appendix B) is the following set of constraints that a quantum agent can satisfy without penalty to their ability to achieve peak memory compression advantage:
\begin{itemize}
\item The agent receives input stimuli $\{x\}$ encoded in the computational basis states $\{\ket{x}\}$.
\item The input stimulus is not consumed by the evolution of the agent; $\Lambda$ preserves the input tape.
\item The agent delivers output actions $\{y\}$ via projective measurements in the computational basis states $\{\ket{y}\}$ of its output tape.
\item The memory states are pure and in one-to-one correspondence with the strategy's causal states $\mathcal{S}$.
\end{itemize}
That is, \emph{generalising beyond these features cannot provide further memory advantage}. With stimuli and actions encoded as classical states, all quantum dynamics occur within the agent's internal dynamics -- the quantum memory advantage is not contingent upon access to a quantum environment. Further, these constraints imply a specific form of memory-minimal quantum agents.

\noindent {\bf Theorem 1:} \emph{A provably memory-minimal quantum agent executing any strategy $\mathcal{P}$ -- for any input strategy $\mathcal{R}$ -- can always be realised using the circuit of \figref{figschematic}(b). That is, the policy $\Lambda$ is realised in two stages. The first stage is a unitary operator $U$ acting on the joint system of (i) the agent's memory $\mathsf{M}$, (ii) input tape containing stimuli $x$ encoded as $\ket{x}$, (iii) output tape initialised in $\ket{0}$, and (iv) `junk' tape also initialised in $\ket{0}$. Then, the output action $y$ is realised by a computational basis measurement of the output tape, and the junk tape is discarded. Moreover, the memory states $\{\ket{\sigma_s}\}$ are all pure and in one-to-one correspondence with the causal states $\mathcal{S}$ of the strategy; the encoding function satisfies Eq.~\eqref{eqcausalencoding}.}

\emph{The unitary evolution can be expressed as}
\begin{equation}
\label{eqgeneral}
U\ket{\sigma_s}\ket{x}\ket{0}\ket{0}=\sum_{y}\sqrt{P(y|x,s)}\ket{\sigma_{\lambda(z,s)}}\ket{x}\ket{y}\ket{\psi(z,s)},
\end{equation}
\emph{where $\ket{\psi(z,s)}$ represents the final state of the junk tape before it is discarded.}

This implies that the only effective degrees of freedom in designing an agent's memory encoding lie in the choice of junk states $\{\ket{\psi(z,s)}\}$, as $U$ and the memory states $\{\ket{\sigma_s}\}$ are then defined implicitly through Eq.~\eqref{eqgeneral}. However, not every choice of junk states is physically-realisible, due to the constraint that $U$ is unitary. Consider the overlap of two memory states $s$ and $s'$, given by $c_{ss'}:=\braket{\sigma_s}{\sigma_{s'}}$. Using the condition $U^{\dagger}U=\mathbb{I}$ and defining $d^{z}_{ss'}:=\braket{\psi(z,s)}{\psi(z,s')}$, from Eq.~\eqref{eqgeneral} we obtain
\begin{equation}
\label{eqoverlaps}
c_{ss'}=\sum_y \sqrt{P(y|x,s)P(y|x,s')}c_{\lambda(z,s)\lambda(z,s')}d^{z}_{ss'}.
\end{equation}
Though expressed for a given stimulus $x$, consistency requires that this equation yield identical $\{c_{ss'}\}$ for all $x$. While this constraint can always trivially be satisfied by setting $\ket{\psi(z,s)}=\ket{s}$ for all $z$, this enforces that quantum memory states are mutually orthogonal, recovering the classical $\varepsilon$-transducer and removing all quantum memory advantage. The crux of the quantum advantage is thus in finding junk states that admit non-orthogonal memory states, and optimising their assignment to maximise it.

It is tempting to look for simple junk states that are just complex scalars, removing the need for junk tape altogether (as the corresponding phase can be absorbed by the output tape). However, this is generally impossible.

\noindent{\bf Theorem 2:} \emph{Junk states $\{\ket{\psi(z,s)}\}$ cannot always be assigned as complex scalars. There exist strategies that can only be executed by quantum agents with access to a multi-dimensional junk tape that is discarded into the environment at each timestep.}

To prove this theorem, let us first suppose that (contrary to the theorem) the junk states are simply a set of complex scalars $\{\exp(i\varphi_{zs})\}$, i.e., \mbox{$d^z_{ss'}=\exp(i(\varphi_{zs'}-\varphi_{zs}))$}. Substituting into Eq.~\eqref{eqoverlaps}, we obtain
\begin{equation}
\label{eqphaseoverlapone}
c_{ss'}=\sum_y\sqrt{P(y|x,s)P(y|x,s')}e^{i(\varphi_{zs'}-\varphi_{zs})}c_{\lambda(z,s)\lambda(z,s')}.
\end{equation} 
The left-hand side of this equation represents the overlaps of the memory states, and so for consistency we must have that the right-hand side is equal for all possible stimuli $x$. To prove the theorem, we need only establish that there is at least one strategy for which no set of phases $\{\varphi_{zs}\}$ exists that can satisfy this condition.

\begin{figure}
\includegraphics[width=\linewidth]{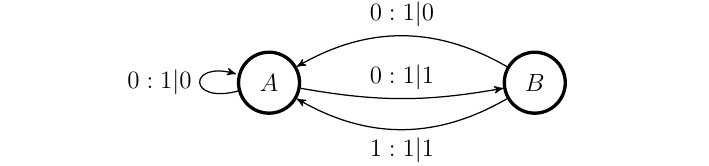}
\caption{{\bf Agents generally need to discard junk information.} Hidden Markov model representation of the example process for Theorem 2. Nodes represent states and edges transitions; the notation $y:p|x$ denotes that on input stimulus $x$ the indicated transition occurs with probability $p$ whilst outputting action $y$. The strategy requires two internal states $\{A,B\}$, two stimuli $\{0,1\}$ and two actions $\{0,1\}$. On stimulus $0$, both states enact $0$ with certainty and transition to state $A$; on stimulus $1$, state $A$ acts with $0$ and transitions to $B$, while state $B$ acts with $1$ and transitions to $A$.}
\label{figdissipate}
\end{figure}

Consider the strategy illustrated in \figref{figdissipate}. For this strategy, Eq.~\eqref{eqoverlaps} demands $c_{AB}=0$ as there is no overlap in future statistics for stimulus 1. Meanwhile, we must then have that $d^{0,0}_{AB}=0$ -- clearly this cannot be satisfied if $\ket{\psi(0,0,A)}$ and $\ket{\psi(0,0,B)}$ differ only by a phase factor. Thus, Theorem 2 is proven.

This is not an isolated example. In Technical Appendix C we derive a sufficiency condition on the strategy that indicates Eq.~\eqref{eqphaseoverlapone} cannot be satisfied for any set of phases $\{\varphi_{zs}\}$, and hence non-scalar junk is required. Informally, this condition holds when the strategy has two states which must give rise to very similar behaviour on one string of possible future stimuli, and very differently on another. The above example represents an extreme case of this. The requirement of non-trivial junk has operational significance, as it mandates that the agent discard information into the environment at each timestep, corresponding to a source of thermal dissipation. The next theorem suggests this dissipation manifests from the data processing inequality.

\noindent {\bf Theorem 3:} \emph{The magnitude of the overlap between any pair of quantum memory states cannot exceed the overlap of their future statistics for any input strategy $\mathcal{R}$;
\begin{equation}
\label{eqbound}
|c_{ss'}|\leq\min_{\mathcal{R}}\sum_{\fut{y}}\sqrt{P(\fut{y}|s,\fut{X})P(\fut{y}|s',\fut{X})}.
\end{equation}}

Physically, this can be understood as requiring that the future statistics do not provide a means of distinguishing between quantum memory states beyond what is information-theoretically possible, imposing a constraint on their maximum fidelity~\cite{suen2017classical}. In Technical Appendix D we show how this bound can be calculated. However, this bound cannot always be saturated; a counterexample is provided in Technical Appendix E.

\section{Systematic quantum agent design}

We now provide a systematic method for assigning junk states such that the corresponding quantum agents achieve superior memory efficiency relative to memory-minimal classical~\cite{barnett2015computational} and prior quantum counterparts alike~\cite{thompson2017using}. The design involves an effective representation of each of the memory states as a tensor-product form $\ket{\sigma_s}=\bigotimes_x \ket{\sigma_s^x}$, where the $\{\ket{\sigma_s^x}\}$ behave as memory states specialised to each input (see Technical Appendix F). These have associated overlaps $c_{ss'}^x:=\braket{\sigma_s^x}{\sigma_{s'}^x}$, such that $c_{ss'}=\prod_x c_{ss'}^x$. In this representation we identify the junk states as $\ket{\psi(z,s)} = \bigotimes_{x'\neq x}\ket{\sigma^{x'}_s}$, and correspondingly, their overlaps (for pairs with identical $z$) as $d^{z}_{ss'}=\prod_{x'\neq x} c_{ss'}^{x'}$. In Technical Appendix F we prove that for any strategy, a unitary of the form Eq.~\eqref{eqgeneral} can always be found that is based on these states. Given a strategy's $\varepsilon$-transducer, Algorithm 1 then provides a systematic means of designing quantum agents with this encoding.
\begin{algorithm}[H]
\caption{\textsf{Systematic quantum agent encoding}}
\begin{flushleft}
\emph{Inputs}: Causal states $\mathcal{S}$, transition probabilities $P(Y|X,S)$, and update rule $\lambda(z,s)$. \\
\emph{Outputs}: Quantum memory states $\{\ket{\sigma_s}\}$, evolution operator $U$.
\end{flushleft}
\begin{algorithmic}[1]
\STATE Construct the set of multivariate polynomial equations
\begin{equation}
\label{eqoverlap}
c^x_{ss'}=\sum_y\sqrt{P(y|x,s)P(y|x,s')}\prod_{x'}c_{\lambda(z,s)\lambda(z,s')}^{x'}
\end{equation}
defined $\forall s,s'\in\mathcal{S},x\in\mathcal{X}$ and solve to obtain $\{c_{ss'}^x\}$.
\STATE Use a reverse Gram-Schmidt procedure~\cite{dennery1996mathematics, binder2018practical} to construct quantum memory states $\{\ket{\sigma_s}\}$ from overlaps \mbox{$c_{ss'}=\prod_x c_{ss'}^x$}, and junk states $\{\ket{\psi(z,s)}\}$ from overlaps  $d^{z}_{ss'}=\prod_{x'\neq x} c_{ss'}^{x'}$.
\STATE Construct the columns of $U$ explicitly defined in Eq.~\eqref{eqgeneral}.
\STATE Fill the remaining columns of $U$, using a Gram-Schmidt procedure to ensure orthogonality with existing columns.
\end{algorithmic}
\end{algorithm}
In this encoding, any given pair of memory states has non-zero overlap iff there is no string of input stimuli for which they are certain to produce distinguishable strings of output actions; provided at least one such pair exists, the quantum agent exhibits a memory advantage over provably minimal classical counterparts~\cite{nielsen2000quantum, gu2012quantum, thompson2017using}. Note that despite their factorised representation presenting as an $|\mathcal{S}||\mathcal{X}|$-dimensional space, the reverse Gram-Schmidt procedure ensures that memory states can be supported by a memory system of at most $|\mathcal{S}|$ dimensions.

\section{Scaling advantage}

The memory advantage of quantum agents can grow without bound. Consider a setting where an agent's optimal strategy depends on tracking some continuous parameter of its environment $\tau$. This can occur when naturally continuous parameters are involved, such as spatial position or time. Alternatively, for strategies with a dependence on events long ago in the past, the set of pasts $\{\past{z}\}$ can be mapped to a continuous parameter over the interval $[0,1)$, by taking $\past{z}$ to specify a $|\mathcal{Z}|$-ary fraction. In either case, small differences in $\tau$ often require only slightly different responses to future stimuli. However, if an agent must store $\tau$ precisely, it requires an unbounded amount of memory. 

To circumvent this, the conventional classical method is to adopt coarse-graining, in which an approximation of the optimal strategy $\mathcal{P}$ is executed based on storing $\tau$ only to some finite precision. That is, $\tau$ is divided up into a set of discrete bins, and all values of $\tau$ within a given bin are mapped to the same memory state. An $n$-bit precision coarse-graining divides $\tau$ into $2^n$ such bins, each of width $\delta \tau^{(n)}$; the corresponding coarse-graining of the strategy is denoted $\mathcal{P}^{(n)}$. For a classical agent, the memory cost then diverges linearly with $n$~\cite{marzen2015informational}, forcing a trade-off between precision and memory cost.

On the other hand, quantum agents may be able to avoid such divergences. Consider a family of quantum agents that implement coarse-grainings $\mathcal{P}^{(n)}$ of a strategy $\mathcal{P}$ at each level of precision $n\in\mathbb{N}$. Consider also the following pair of convergence conditions, defined formally in Technical Appendix G:
\begin{itemize}
\item {\bf Distributional convergence:} The steady-state probability (densities) of the memory states converge exponentially with increasing precision.
\item {\bf Memory-overlap convergence:} The overlaps of each pair of memory states converge exponentially with increasing precision.
\end{itemize}
These convergence conditions encapsulate the intuition that if the strategy varies smoothly with a continuous parameter, then so too may the properties of the memory states of a quantum agent executing the strategy. When these conditions are met, \emph{a quantum agent can execute the strategy $\mathcal{P}$ to arbitrary precision with bounded memory cost, giving rise to a scaling advantage over classical agents}. The formal statement of this result is given in Theorem 4, which may be found in Technical Appendix G together with its proof.

\begin{figure}
\includegraphics[width=\linewidth]{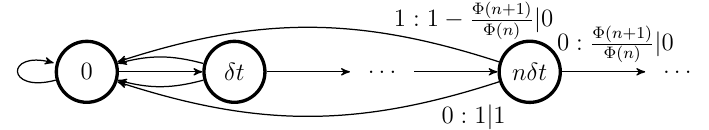}
\includegraphics[width=\linewidth]{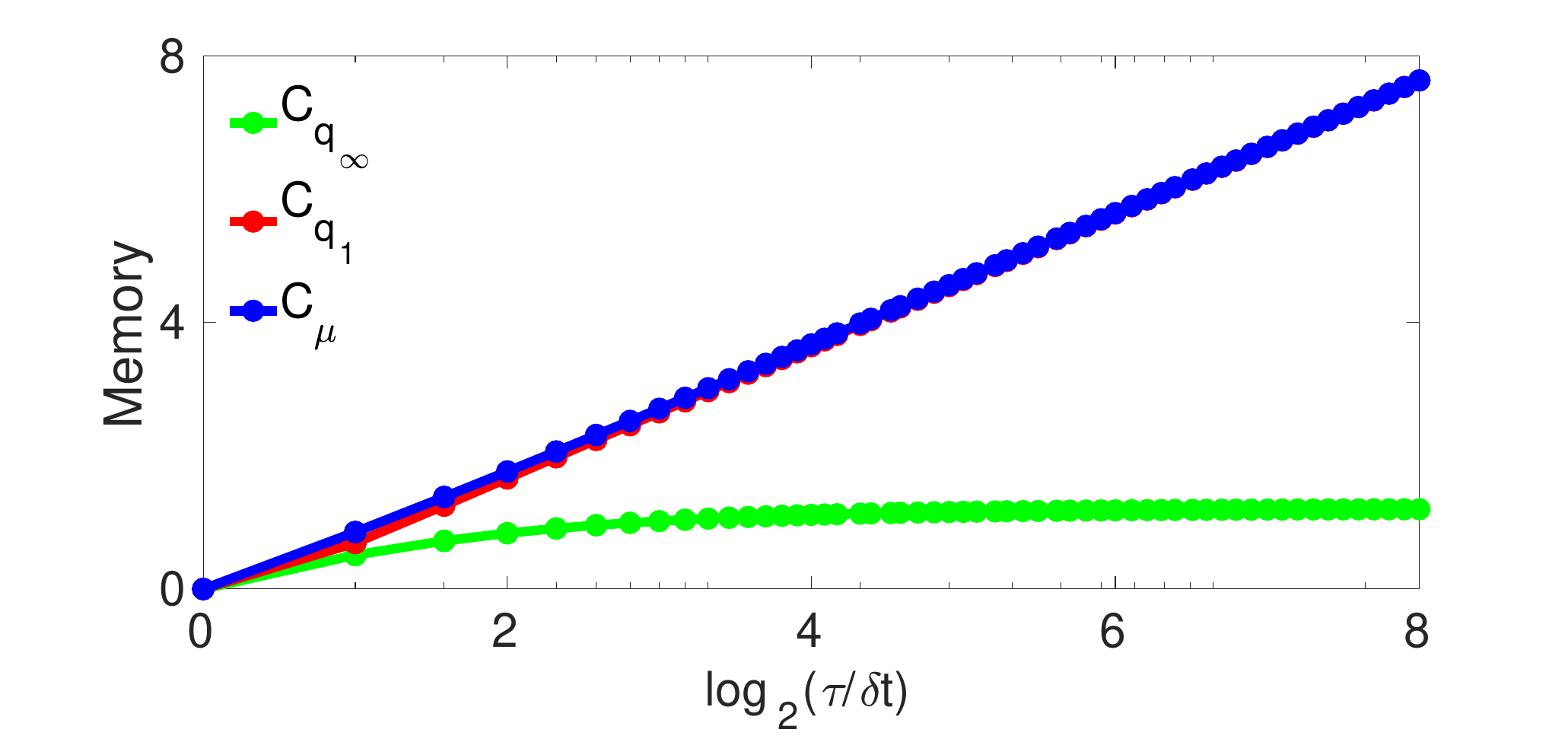}
\caption{{\bf Unbounded scaling advantage.} (a) Hidden Markov model representation of resettable stochastic clocks; state labels represent the number of timesteps since the last event; $\Phi(n):=\int_{n\delta t}^\infty\phi(t)dt$ is the survival probability. (b) Associated memory costs for executing such a strategy, showing the advantage of our quantum agent ($C_{q_\infty}$) growing unbounded relative to previous quantum ($C_{q_1}$) and minimal classical agents ($C_\mu$) with refinement of timesteps.}
\label{figexample}
\end{figure}

We illustrate an example of such scaling advantages occurring for agents tasked with executing certain strategies requiring co-ordinated stimuli-action responses over an increasingly greater number of timesteps. We demonstrate this with an example family of resettable stochastic clocks. In this setting, the agent is tasked with behaving as a clock with stochastic tick events, that may be reset by an external stimulus. This stimulus can take two values: $x=0$ for `evolve normally', and $x=1$ for `reset', while possible actions are $y=0$ for `no tick' and $y=1$ for `tick'. When $x=0$, the agent behaves as a stochastic clock~\cite{woods2018quantum, yang2020ultimate}, modelled by a renewal process~\cite{smith1958renewal} where the agent emits a tick at stochastic intervals $t$ governed by a distribution $\phi(t)$. Upon receiving $x=1$ however, the agent must immediately reset its time-counter, such that the clock behaves as though it has just ticked. The agent must replicate this behaviour to some desired temporal resolution, such that time is broken into finite timesteps $\delta t$ -- as illustrated diagrammatically in \figref{figexample}(a), with further details in Technical Appendix H. For a given $\phi(t)$ this prescribes a family of coarse-grained strategies parameterised by $\delta t$.

In Technical Appendix H we show that our quantum agents satisfy the convergence conditions for a large class of $\phi(t)$ representing typical resettable stochastic clocks, and thus may execute them to arbitrary precision with a bounded cost. Meanwhile, the memory-minimal classical models must store an ever-increasing amount of information as $\delta t$ is refined. That is, \emph{our quantum agents converge to a finite memory cost, while the classical agents diverge}. \figref{figexample}(b) highlights this by comparing the scaling of our quantum agents (labelled $C_{q_\infty}$) with the memory-minimal classical ($C_\mu$) and best prior quantum counterparts ($C_{q_1}$) for the particular case where $\phi(t)$ is uniformly distributed over the interval $[0,\tau]$, and resets are triggered at a constant rate $1/2\tau$~\footnote{Note that though we are considering the memory costs with respect to a particular input strategy, we drop the $\mathcal{R}$ from the subscript for notational brevity.}.

\section{Discussion}

We have introduced a general framework for adaptive agents that can capitalise on access to a quantum memory to reduce the information they must track about past stimuli and actions.
Key to this, we isolated the features of an agent that are relevant to memory advantages, and showed that they are in direct correspondence with the information it discards into its environment. Coupled with this, we provided a systematic algorithm for encoding the memory states of a quantum agent for any strategy, achieving a memory advantage relative to minimal classical and prior state-of-the-art quantum counterparts. Moreover, this advantage can grow without bound. These advantages can be utilised by agents for both executing fixed strategies and in running candidate strategies during their development~\cite{kaelbling1996reinforcement, bu2008comprehensive, kober2013reinforcement, sutton2018reinforcement}, as well as by researchers modelling the behaviour of agents. Our systematic quantum agent design may also be used for enhancing mechanical agents, for example, by endowing smart technologies with quantum processors. Our framework is agnostic to the specific engineering details of its implementation, and so can be realised with any quantum architecture that can receive (classical) input, and process and store quantum information according to the required policy evolution of Eq.~\eqref{eqgeneral}. Proof-of-principle demonstrations are feasible with current setups, by, for example, adapting prior implementations of quantum models of passive stochastic processes in photonic setups~\cite{ghafari2018single, ghafari2019interfering} to undergo different evolutions at each timestep conditional on the input.

Our results use entropic benchmarks for the memory, thus naturally assuming an ensemble setting. They describe quantum memory advantages with operational relevance for multiple agents implementing a strategy in parallel with shared memory~\cite{elliott2018superior}. This aligns well with scenarios where one wishes to sample over the conditional distributions for various strategies, for example in Markov Chain Monte Carlo-type methods~\cite{gilks1995markov}. A compelling extension is to single-shot settings, where one may instead consider the max entropy -- the dimension of the state-space inhabited by the memory states. Single-shot advantages have been found for quantum models of passive stochastic processes~\cite{thompson2018causal, loomis2019strong, ghafari2018single, liu2019optimal, elliott2020extreme, elliott2021quantum}, and for specific cases of input-output behaviour modelling repeated measurement of a quantum system~\cite{cabello2016thermodynamical}. Since our general treatment ultimately relates to what can affect memory state overlaps, many of our results will continue to hold in single-shot settings -- in particular our form of the memory-minimal quantum agent -- and thus can direct the search for systematic encodings based on other such benchmarks. Based on links established between quantum compression advantages and thermal efficiency in stochastic modelling~\cite{loomis2020thermal, elliott2021memory}, one may expect that our quantum agents are also able to execute their strategies with less thermal dissipation than classical counterparts.

A further enticing extension would be to the case where only near-faithful execution of the strategy is required -- that is, some error is tolerated~\cite{ho2020robust, yang2020measures, elliott2021quantum}. Our quantum agents bear a similarity to models of quantum walks with memory~\cite{flitney2004quantum, mcgettrick2009one, rohde2013quantum} and other instances of memory compression through quantum processing~\cite{rozema2014quantum, yang2018quantum} such as quantum auto-encoders~\cite{romero2017quantum, wan2017quantum, pepper2019experimental, huang2019realization}. Moreover, our general form for quantum adaptive agents Eq.~\eqref{eqgeneral} produces superpositions of all possible future trajectories for the input~\cite{binder2018practical}, potentially allowing for interference experiments that probe the overlap in the distributions of different strategies~\cite{ghafari2019interfering}, or different input sequences. One can also consider superpositions of input sequences, akin to algorithms in quantum-enhanced reinforcement learning~\cite{paparo2014quantum, dunjko2016quantum, dunjko2018machine}, where our agents may augment existing quantum speed-ups with extreme memory advantages.

\section*{Technical Appendix}

\subsection*{A: Framework (Extended)}

Here we provide further details of the framework used to describe adaptive agents, containing additional material relevant to the remaining appendices. We begin by formally defining an adaptive agent, as introduced in the main text.

\noindent {\bf Definition 1:} (Adaptive agents) \emph{An adaptive agent is defined by the tuple $(\mathcal{X},\mathcal{Y},\{\sigma_m\},f,\Lambda)$, where
\begin{itemize}
\item $\mathcal{X}$ is the set of stimuli the agent can recognise;
\item $\mathcal{Y}$ is the set of actions the agent can perform;
\item $\{\sigma_m\}$ is the set of memory states the agent can store in its memory system $\mathsf{M}$, labelled by an index \mbox{$m\in\mathcal{M}$};
\item $f:\past{\mathcal{Z}}\to\{\sigma_m\}$ is the encoding function that determines the memory state to which the agent assigns each history $\past{z}$;
\item $\Lambda:\mathcal{X}\times\{\sigma_m\}\to\mathcal{Y}\times\{\sigma_m\}$ is the agent's policy, describing how the agent selects action $y$ in response to stimulus $x$ given its current memory state, and how the memory state is updated.
\end{itemize}}

An encoding $f$ is said to be a valid encoding of a strategy $\mathcal{P}$ if there exists a policy $\Lambda$ by which the agent is able to execute actions in a manner statistically faithful to the strategy for every possible history and sequence of future stimuli. That is, $f$ is valid iff \mbox{$\exists\Lambda:\left(P(\fut{Y}|m,\fut{x})=P(\fut{Y}|\past{z},\fut{x})|f(\past{z})=\sigma_m\right)\forall \past{z},\fut{x}$}. An agent with such a policy and encoding function is then said to faithfully execute strategy $\mathcal{P}$. Hereon, we consider such faithful agents. The physics of the memory states determines the physics of the agent; that is, a classical agent can only store classical states in its memory and use classical dynamics for its policy, while for a quantum agent $\mathsf{M}$ can support quantum states, and $\Lambda$ takes the form of a quantum channel.

A strategy $\mathcal{P}$ can be described as a conditional distribution $P(Y|\past{Z},X)$. Mathematically, this corresponds to a stochastic input-output process~\cite{khintchine1934korrelationstheorie, crutchfield1994calculi, barnett2015computational, thompson2017using}, where the stimuli are the inputs, and the actions the outputs, and the process maps stimuli and past actions to future actions. Consequently, our results encompass as a special case quantum models of passive stochastic processes -- stochastic processes that evolve autonomously without environmental input -- by taking the input alphabet to consist only of a single symbol (i.e., the strategy does not condition on any observed stimuli). 

There are certain conditions implicitly placed on these input-output processes due to the limits of what an agent can predict about the future. That is, an agent cannot leverage information about future events that cannot be deduced from what they have already seen. The two conditions are referred to as the agent being non-anticipatory and causal~\cite{barnett2015computational}. The former requires that the strategy for choosing the current action must not depend on future input stimuli whenever these future stimuli are generated independently of past actions, i.e., $P(Y|\past{Z},X)=P(Y|\past{Z},\fut{X})$~\cite{caruso2014quantum, barnett2015computational, thompson2017using}. The latter requires that the memory of an agent can depend only on the past, and not the future -- i.e., that $f$ is a deterministic map from histories to memory states~\cite{thompson2018causal}. We also assume that the strategy is stationary (time-invariant), such that the weightings $P(Y|\past{Z},X)$ are independent of the timestep $t$. Pasts and futures are taken to consist of semi-infinite strings of stimuli and actions. That is, at $t=0$ we take $\past{x}:=\lim_{l\to\infty}x_{-l:0}$ and $\fut{x}:=\lim_{l\to\infty}x_{0:l}$, where  $x_{k:l}:=x_k,x_{k+1},\ldots,x_{l-1}$ denotes a contiguous string in the interval $k\leq t<l$.

In the main text we note that the input stimuli are in full generality drawn from an input strategy, where the stimuli manifest as actions of the agent's environment, potentially conditioned on the previous actions of the agent. 

\noindent {\bf Definition 2:} (Input strategies) \emph{An input strategy $\mathcal{R}$ is an input-output stochastic process specified by a conditional distribution $R(X_t|\past{Z}_t)$ used to generate input stimuli of an adaptive agent. That is, it maps histories $\{\past{z}\}$ to the next stimulus received by the agent.}

The subscripts indicate that the input strategy can have a temporal dependence (i.e., that it need not be stationary), while the conditioning on the entire history allows the stimuli to have a dependence on the actions of the agent. In the case where the stimuli are generated independent of the agent's actions, $\mathcal{R}$ reduces to a passive stochastic process. Note that in previous works the worst-case memory cost was considered only with respect to such input stochastic processes~\cite{barnett2015computational, thompson2017using}, rather than the more general input strategies described here.

\subsection*{B: Proof of Theorem 1}

We begin with the most general form a quantum adaptive agent can take, progressively examining each aspect to ascertain whether it is essential to its function, and whether it offers potential compression advantages -- in order to constrain to the most general functional agent. 

In full generality, at each timestep, we have an evolution (i.e., a quantum channel) that acts on the current memory state $\rho_m$ and the input stimulus $x$, encoded into a state $\rho_x$. These are mapped by the policy to an output action $y$, extractable from a state $\rho_Y(x,m)$ with probability $P(y|x,m)$, and an updated memory state $\rho_{m'}$ according to $m'=\lambda(x,y,m)$. For a complete accounting, we allow for the inclusion of a `blank' ancilla $\ket{0}$ tape with the input, and a `junk' state $\ket{\psi(x,y,m)}$ with the output -- both may without loss of generality be considered in their purified form~\cite{nielsen2000quantum}. 

\noindent {\bf Lemma 1:} \emph{There is no further quantum advantage from allowing memory states to be non-pure. Moreover, there is no further advantage for the memory states to be anything other than in one-to-one correspondence with the causal states of the $\varepsilon$-transducer.}

These results follow by generalising the so-called causal state correspondence~\cite{suen2017classical} and mixed state exclusion~\cite{thompson2018causal} found for quantum models of passive stochastic processes to the case of strategies. These establish that the memory states of the minimal quantum agents are in one-to-one correspondence with the causal states of the strategy, and can be instantiated as pure states. Our proofs of the generalisations largely follow those of the originals, with the modification to input-conditioned probability distributions.

\noindent {\bf Proposition 1:} (Causal state correspondence) \emph{For any strategy $\mathcal{P}$ with causal encoding function $f_\varepsilon$, there exists a memory-minimal causal, non-anticipatory quantum agent implementing the strategy with memory encoding function $f$ that satisfies}
\begin{equation}
f_\varepsilon(\past{z})=f_\varepsilon(\past{z}')\Leftrightarrow f(\past{z})=f(\past{z}')
\end{equation}
\emph{for all past histories $\past{z}$ and $\past{z}'$.}

We first prove the reverse direction through its contrapositive. Suppose we had two histories $\past{z}$ and $\past{z}'$ belonging to different causal states, but mapped to the same memory state by $f$. The former condition implies $P(\fut{Y}|\past{z},\fut{X})\neq P(\fut{Y}|\past{z}',\fut{X})$, while the the second implies $f(\past{z})=f(\past{z}')$. Since the two memory states are identical, there is no quantum operation that could distinguish between them, and hence no operation that could produce different future statistics from them -- and thus no quantum agent can generate the correct conditional future statistics for both histories. Therefore, we require $f(\past{z})\neq f(\past{z}')$ if $f_\varepsilon(\past{z})\neq f_\varepsilon(\past{z}')$.

The forward direction follows from concavity of entropy~\cite{nielsen2000quantum}. Consider the set of histories $\{\past{z}\}$ belonging to causal state $s$. We define the contribution to the steady-state of the memory coming from histories not in this set as $\rho_{\bar{s}}=\sum_{\past{z}\notin s}P(\past{z})f(\past{z})$, and hence we can express $\rho=\sum_{\past{z}\in s}P'(\past{z})[P(s)f(\past{z})+\rho_{\bar{s}}]$, where $P'(\past{z})=P(\past{z})/P(s)$. From the concavity of entropy, it follows that
\begin{align}
S_{\mathrm{vN}}[\rho]&\geq \sum_{\past{z}\in s}P'(\past{z})S_{\mathrm{vN}}[P(s)f(\past{z})+\rho_{\bar{s}}]\nonumber\\
&\geq\min_{\past{z}\in s}S_{\mathrm{vN}}[P(s)f(\past{z})+\rho_{\bar{s}}].
\end{align}
Let $\past{z}^*$ be the particular history that minimises this inequality. We thus have that for any valid quantum agent, an encoding which assigns all histories belonging to $s$ to $f(\past{z}^*)$ will have lower or equal entropy. Moreover, the modified encoding is also a valid encoding: as the future statistics the agent must produce from $f(\past{z})$ for any other history $\past{z}'\in s$ are the same as those that must be produced from $f(\past{z}^*)$, an encoding with $f(\past{z})=f(\past{z}^*)\forall \past{z}\in s$ will produce the correct future statistics. This procedure can be repeated for histories belonging to all other $s'\neq s$, and we hence find that for any quantum agent there exists another quantum agent implementing the same strategy with lower or equal entropy using an encoding function that assigns all histories in the same causal state to the same memory state.

\noindent {\bf Proposition 2:} (Mixed state exclusion) \emph{For any quantum agent implementing a strategy $\mathcal{P}$ using a valid encoding with memory states $\{\rho_m\}$, there exists a valid encoding of lower or equal entropy with pure memory states $\{\ket{\sigma_m}\}$.}

We start by invoking the causal state correspondence, such that our goal is to show that for any valid quantum encoding for a strategy $\mathcal{P}$ with memory states $\{\rho_s\}$, there exists a valid quantum encoding of lower or equal entropy with pure memory states $\{\ket{\sigma_s}\}$. Suppose a particular memory state $\rho_s$ is non-pure, such that we can decompose it as $\rho_s=\sum_jp_j\ket{a_j}\bra{a_j}$ for some set of pure states $\{\ket{a_j}\}$. Recall that causality demands the memory contain no information about the future that cannot be determined from its past; given the past stimuli and actions, there must be no correlations between the memory states and the futures they produce. This means that each of the states $\{\ket{a_j}\}$ in our decomposition of $\rho_s$ must all individually give rise to the same statistical futures as $\rho_s$, and thus a valid quantum encoding can be formed by replacing $\rho_s$ with any of the $\ket{a_j}$. We again collect all contributions to the steady-state from terms belonging to causal states other than $s$ as $\rho_{\bar{s}}$, such that $\rho=\sum_j p_j(P(s)\ket{a_j}\bra{a_j}+\rho_{\bar{s}})$. From concavity of entropy:
\begin{align}
S_{\mathrm{vN}}[\rho]&\geq\sum_j p_j S_{\mathrm{vN}}\left[P(s)\ket{a_j}\bra{a_j}+\rho_{\bar{s}}\right]\nonumber\\
&\geq\min_j S_{\mathrm{vN}}\left[P(s)\ket{a_j}\bra{a_j}+\rho_{\bar{s}}\right].
\end{align}
Let $\ket{a_j}$ be the particular state that minimises the inequality, and designate it as $\ket{\sigma_s}$. We can thus obtain a valid quantum encoding of lower or equal entropy after replacing $\rho_s$ with $\ket{\sigma_s}$. We can repeat the procedure for the memory states corresponding to other causal states, thus obtaining a valid encoding of lower or equal entropy where all memory states are pure.

Note that the above lemma is not specific to the von Neumann entropy, and holds for \emph{any} entropy satisfying concavity. Since Lemma 1 allows us to restrict our attention to pure memory states, and the Gram matrix representation~\cite{horn1990matrix} of an ensemble of pure quantum states allows us to express the entropy as a function of pairwise overlaps of the states, we can hereon consider features of the agent that can affect the overlap of memory states to be synonymous with those that can (potentially) reduce the memory cost.

\noindent {\bf Lemma 2:} \emph{There is no memory advantage to encoding the input stimulus $x$ as anything other than the computational basis state $\ket{x}$. Moreover, the input state need not be consumed by the evolution.}

Consider that for each input state $\rho_x$ there is a computational basis state $\ket{x}$ appended to it which remains unchanged by the evolution. Then, since it can be factored out it can be seen that it does not influence the overlaps of the memory states, and hence does not affect the amount of information stored. However, we can perform operations conditioned on the appended state, which, since they are orthogonal, allows us to imprint the $\rho_x$ directly onto part of the blank ancilla space and proceed as before. Specifically, we can realise this as a unitary operation $U_\mathcal{X}\ket{x}\ket{0}\ket{0}$, where the third subspace is discarded into the junk and $\rho_x$ is the resulting state of the second subspace after tracing out the other two. We see that it is sufficient to consider orthogonal input states $\{\ket{x}\}$, which can be used to mimic the effect of any set of input states -- in effect, accounting for the pre-processing used to create $\rho_x$ from the input stimulus as part of the evolution. As the appended input space is not affected by the evolution, it can be later used to retrieve the input stimulus.

\noindent {\bf Lemma 3:} \emph{There is no memory advantage for the extraction of $y$ from $\rho_Y(x,m)$ to be anything other than a projective measurement in the computational basis.}

The output action must be extracted from $\rho_Y(x,m)$ through measurement. Neumark's dilation theorem allows us to express any quantum measurement as a projective measurement on a purified state in a larger space~\cite{neumark1943spectral, stinespring1955positive, akhiezer2013theory} -- we can consider any model of the extraction that does not strictly use projective measurements to effectively be relegating this extended space into the junk. This dilation does not change the evolution of the memory state and hence there is no penalty to working with the projective measurement picture. As Lemma 2 allows us to take the input states $\ket{x}$ to be orthogonal we can consider the output subspace to always be conditionally rotated at the end of the evolution such that the appropriate measurement basis is the computational basis, independent of the input stimulus.

With these lemmas, we can express the evolution at each timestep by a global unitary operator $U$ [Eq.~\eqref{eqgeneral}]. The amplitudes follow from the requirement that outcome $y$ must be obtained with probability $P(y|x,s)$~\cite{binder2018practical, liu2019optimal}, and without loss of generality can be taken to be real by offloading any phase factor into the junk subspace. 

\subsection*{C: Sufficiency condition for necessity of junk}

Here we provide a sufficient (but not necessary) condition on a strategy upon which no physically-realisable quantum agent can implement said strategy without use of discarded junk states. 

For a given strategy, consider a pair of states $s$ and $s'$ and strings of stimuli $x_{0:L}$ and actions $y_{0:L}$ for which $\lambda(z_{0:L},s)=\lambda(z_{0:L},s')$, where the output of the update function on a string of stimuli/actions is understood to be the sequential application of the update for each timestep (i.e., $\lambda(z_0z_1,s):=\lambda(z_1,\lambda(z_0,s))$. Let us for shorthand denote $p:=P(y_{0:L}|x_{0:L},s)$ and $p':=P(y_{0:L}|x_{0:L},s')$. Iterating through Eq.~\eqref{eqphaseoverlapone}, we obtain that this provides a contribution of magnitude $\sqrt{pp'}$ to the overlap of the two states if there is no junk. The magnitude of the remaining terms (corresponding to other action strings) must then be bounded by $\sqrt{(1-p)(1-p')}$. If $p+p'=1+\alpha$ for some non-negative $\alpha$, we then have that
\begin{equation}
|c_{ss'}|\geq\sqrt{pp'}-\sqrt{(1-p)(1-p')}>\frac{\alpha}{2}.
\end{equation}
This can be verified by direct substitution into the condition $\sqrt{pp'}-\sqrt{(1-p)(1-p')}>\alpha/2$ after rearrangement and squaring, and using that $p+p'>1$ implies $(1-p)(1-p')<p(1-p)<1/4$.

Now consider another string of stimuli $x_{0:L}'$. Iterating Eq.~\eqref{eqphaseoverlapone}, we have that $|c_{ss'}|\leq\sum_{y_{0:L}}\sqrt{P(y_{0:L}|x_{0:L}',s)P(y_{0:L}|x_{0:L}',s')}$, placing an upper bound on the permissible overlap of the states. If this is less than $\alpha/2$, then a junk-free encoding cannot reduce the overlap as mandated by the $x_{0:L}$ string of stimuli sufficiently low enough to reach this bound. In such circumstances it is then necessary to have the dissipation of junk states that allow further reductions in the overlaps.

To state this sufficiency condition more directly: If a strategy has a pair of states $s$ and $s'$ for which their exists a pair of stimuli-action strings $x_{0:L}$ and $y_{0:L}$ satisfying $\lambda(z_{0:L},s)=\lambda(z_{0:L},s')$ and $P(y_{0:L}|x_{0:L},s)+P(y_{0:L}|x_{0:L},s')=1+\alpha$ for some non-negative $\alpha$, and another stimuli string $x_{0:L}'$ satisfying $\sum_{y_{0:L}}\sqrt{P(y_{0:L}|x_{0:L}',s)P(y_{0:L}|x_{0:L}',s')}\leq\alpha/2$, then the strategy cannot be implemented by an agent that does not utilise junk states.

We note that this condition need only be met for a single pair of stimuli strings on a single pair of states in order for the agent to require junk.

\subsection*{D: Bounding quantum memory state overlaps}

Eq.~\eqref{eqbound} in the main text places an upper bound on the overlap between any pair of quantum memory states, based on the distinguishability of their future statistics. Here, we provide two methods by which this bound can be calculated: the first method is approximate, with a computational cost that grows quadratically with the number of causal states and linearly with the depth of the approximation; the second is exact, but bears an exponential scaling in cost.

Suppose we are told that the memory has been initialised in one of two memory states $\{\ket{\sigma_s},\ket{\sigma_{s'}}\}$, and we are asked to determine which one with a fixed number of input stimuli $L$. Obviously, if $L=0$, we are unable to distinguish between the possible states. With $L=1$, we wish to choose the stimulus $x$ that minimises the fidelity of the next output action, i.e., \mbox{$\mathrm{argmin}_x\sum_y\sqrt{P(y|s,x)P(y|s',x)}$.} For $L=2$, we are able to choose the second stimulus based on the action output in response to the first, and the first stimulus should be chosen bearing this in mind. Denoting $F_{ss'}^{(1)}:=\min_x\sum_y\sqrt{P(y|s,x)P(y|s',x)}$, we see that the best strategy for choosing the first stimulus $x$ is \mbox{$\mathrm{argmin}_x\sum_y\sqrt{P(y|s,x)P(y|s',x)}F^{(1)}_{\lambda(z,s)\lambda(z,s')}$.} An iterative strategy can be developed, leading to our first method: define $F^{(0)}_{ss'}=1$ and $F^{(L+1)}_{ss'}=\min_x\sum_y\sqrt{P(y|s,x)P(y|s',x)}F^{(L)}_{\lambda(z,s)\lambda(z,s')}$ for all pairs of causal states $\{s,s'\}$; iterate through to the desired precision. Accounting for the symmetries $F^{(L)}_{ss'}=F^{(L)}_{s's}$ and $F^{(L)}_{ss}=1$, at each step there are $|\mathcal{S}|(|\mathcal{S}|-1)/2$ functions to minimise over $|\mathcal{X}|$ arguments each, leading to a scaling cost of $L|\mathcal{X}||\mathcal{S}|(|\mathcal{S}|-1)/2$. It is intuitively clear that allowing for longer input strings cannot decrease the ability to distinguish between the states, and indeed, $F^{(L+1)}_{ss'}\leq F^{(L)}_{ss'}$. The method thus overestimates the upper bound, and as $L\to\infty$, the estimate converges on the bound.

The second method makes use of the fact that for each pair of memory states there is an optimal choice of next input stimulus, conditional on the number of subsequent input stimuli we are able to make. Observing that in the above iterative procedure we should have $F^{(\infty)}_{ss'}=\min_x\sum_y\sqrt{P(y|s,x)P(y|s',x)}F^{(\infty)}_{\lambda(z,s)\lambda(z,s')}$, we can postulate the optimal stimulus for each pair, and solve the associated linear equations. Minimising this over all possible postulates for the optimal stimuli, we obtain the actual bound. However, there are $|\mathcal{X}|^{|\mathcal{S}|(|\mathcal{S}|-1)/2}$ possible assignments of stimuli, and hence the computational cost of this method scales exponentially with the number of causal states.

We can also consider a hybrid of the two methods, to obtain an improved estimate over the first: begin by carrying out the first method to some desired depth $L$, then using the corresponding arguments that minimise the expressions as the postulate, evaluate the recursion relations from the second method.

\subsection*{E: Counterexample to fidelity bound tightness} 

As noted in the main text, counterexamples to the tightness of the fidelity upper bound on memory state overlap Eq.~\eqref{eqbound} exist. Here we provide such a counterexample.

\begin{figure}
\includegraphics[width=\linewidth]{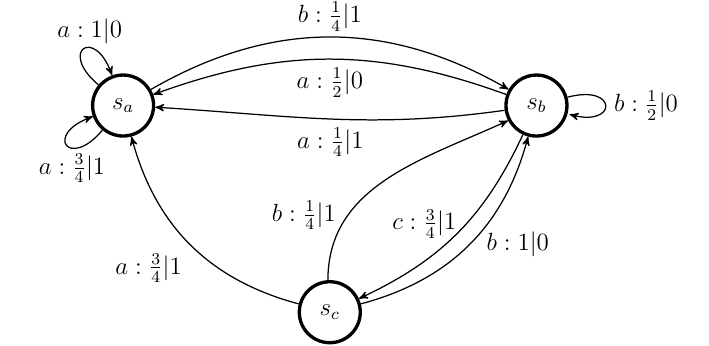}
\caption{{\bf Non-tightness of fidelity bound.} Hidden Markov model representation of a counterexample strategy to the tightness of the fidelity bound, as described in Technical Appendix E.}
\label{figfidelity}
\end{figure}

Consider an agent with three memory states $\{s_a,s_b,s_c\}$, three actions $\{a,b,c\}$ and two stimuli $\{0,1\}$. The dynamic is Markovian, such that after action $y$ the memory transitions to state $s_y$. Let the corresponding strategy be defined by the following probabilities (illustrated in \figref{figfidelity}):
\begin{equation}
\def\arraystretch{1.5}
\begin{array}{c c c}
P(a|0,s_a)=1 & \quad\quad & P(a|1,s_a)=\frac{3}{4}\\
P(a|0,s_b)=\frac{1}{2}& & P(b|1,s_a)=\frac{1}{4}\\
P(b|0,s_b)=\frac{1}{2}& & P(a|1,s_b)=\frac{1}{4}\\
P(b|0,s_c)=1& & P(c|1,s_b)=\frac{3}{4}\\
& & P(a|1,s_c)=\frac{3}{4}\\
& & P(b|1,s_c)=\frac{1}{4}
\end{array}
\end{equation}
with the remaining unspecified probabilities all zero. Each of the states possess non-equal output responses to the stimuli, and so form the causal states of the strategy.

From stimulus 0 we obtain the following upper bounds on memory state overlaps:
\begin{equation}
|c_{ab}|\leq\frac{1}{\sqrt{2}}\quad\quad |c_{bc}|\leq\frac{1}{\sqrt{2}}\quad\quad |c_{ac}|=0,
\end{equation}
while stimlus 1 yields the bounds
\begin{equation}
|c_{ab}|\leq\frac{\sqrt{3}}{4}\quad\quad |c_{bc}|\leq\frac{\sqrt{3}}{4}\quad\quad |c_{ac}|\leq1.
\end{equation}
If the fidelity bound is to be saturated, we must have 
\begin{equation}
|c_{ab}|=\frac{\sqrt{3}}{4}\quad\quad |c_{bc}|=\frac{\sqrt{3}}{4}\quad\quad |c_{ac}|=0.
\end{equation}

For stimulus 1 the evolution must be of the form
\begin{align}
U\ket{\sigma_a}\ket{1}\ket{0}\ket{0}&=\frac{\sqrt{3}}{2}\ket{\sigma_a}\ket{1}\ket{a}\ket{\psi(1,a,s_a)}\nonumber\\&+\frac{1}{2}\ket{\sigma_b}\ket{1}\ket{b}\ket{\psi(1,b,s_a)}\nonumber\\
U\ket{\sigma_b}\ket{1}\ket{0}\ket{0}&=\frac{1}{2}\ket{\sigma_a}\ket{a}\ket{1}\ket{\psi(1,a,s_b)}\nonumber\\&+\frac{\sqrt{3}}{2}\ket{\sigma_c}\ket{1}\ket{c}\ket{\psi(1,c,s_b)}\nonumber\\
U\ket{\sigma_c}\ket{1}\ket{0}\ket{0}&=\frac{\sqrt{3}}{2}\ket{\sigma_a}\ket{1}\ket{a}\ket{\psi(1,a,s_c)}\nonumber\\&+\frac{1}{2}\ket{\sigma_b}\ket{1}\ket{b}\ket{\psi(1,b,s_c)}.
\end{align}
To attain the prescribed values of $|c_{ab}|$ and $|c_{bc}|$ we must have $\ket{\psi(1,a,s_a)}=\exp(i\varphi_1)\ket{\psi(1,a,s_b)}=\exp(i\varphi_2)\ket{\psi(1,a,s_c)}$ -- i.e., equal up to phase factors. However, the condition on $|c_{ac}|$ would then require
\begin{equation}
\frac{3}{4}e^{-i\varphi_2}+\frac{1}{4}\braket{\psi(1,b,s_c)}{\psi(1,b,s_a)}=0,
\end{equation}
which clearly cannot be satisfied. Thus, the fidelity bound cannot be tightly satisfied.

Interestingly, this manifests only for non-trivial strategies; for passive stochastic processes it is always possible to construct a quantum model of the process with trivial (i.e., one-dimensional) junk states that saturates the fidelity bound~\cite{binder2018practical, liu2019optimal}.

\subsection*{F: Details for systematic quantum agent design}

Recall that our systematic encoding is based on a representation using factorised quantum memory states $\{\ket{\sigma_s}\}=\{\bigotimes_x \ket{\sigma_s^x}\}$, with overlaps $c_{ss'}^x:=\braket{\sigma_s^x}{\sigma_{s'}^x}$ and $c_{ss'}=\prod_x c_{ss'}^x$. Consider input stimuli-specific unitaries $\{U_x\}$ that act in the following manner~\cite{binder2018practical, liu2019optimal} on the corresponding input-specialised memory substates:
\begin{equation}
U_x\ket{\sigma_s^x}\ket{0}\ket{0}=\sum_y\sqrt{P(y|x,s)}\ket{\sigma_{\lambda(z,s)}}\ket{y},
\end{equation}
where we have combined the first and second subspaces on the left-hand side together on the right; this implicitly defines the memory substates. We also define a selection operation $U_{\mathrm{select}}$ that ensures that the correct memory state is acted on with the correct $U_x$, conditioned on the input state. Specifically, we define this operation to permute the memory substates conditioned on stimulus $x$ such that the $x$th memory substate is in the first position and exchange the remaining memory substates with the junk. We then act with $U_x$ conditioned on the input state. Defining $U=(\sum_xU_x\otimes\ket{x}\bra{x}\otimes \mathbb{I})U_{\mathrm{select}}$, we obtain the total evolution consistent with Eq.~\eqref{eqgeneral}.

In this representation, the junk states are given by $\ket{\psi(z,s)}=\bigotimes_{x'\neq x}\ket{\sigma^{x'}_s}$, i.e., the unused memory substates corresponding to other input stimuli. Using that $U^{\dagger}U=\mathbb{I}$ we obtain
\begin{equation}
c_{ss'}=\sum_y\sqrt{P(y|x,s)P(y|x,s')}c_{\lambda(z,s)\lambda(z,s')}\prod_{x'\neq x}c_{ss'}^{x'},
\end{equation}
This can then be reduced to be purely in terms of the substate overlaps, recovering Eq.~\eqref{eqoverlap}:
\begin{equation}
c^x_{ss'}=\sum_y\sqrt{P(y|x,s)P(y|x,s')}\prod_{x'}c_{\lambda(z,s)\lambda(z,s')}^{x'}.
\end{equation}
As described in the algorithm, the overlaps can then be found by solving this set of multivariate polynomial equations. A solution always exists for any process that asymptotically synchronises (i.e., \mbox{$\lim_{L\to\infty}H(S_0|Z_{0:L})=0$}): since a sufficiently long string of past stimuli-action pairs allows the causal state to be determined with certainty, by iterating through the recursion relations we obtain the solution
\begin{align}
\label{eqoverlapexpand}
c_{ss'}&\!=\!\!\lim_{L\to\infty}\prod_{x_0}\!\sum_{y_0}\!\!\sqrt{P(y_0|x_0,s)P(y_0|x_0,s')}\nonumber\\
&\!\!\!\!\times\!\prod_{x_1}\!\sum_{y_1}\!\!\sqrt{P(y_1|x_1,\lambda(z_0,s))P(y_1|x_1,\lambda(z_0,s'))}\nonumber\\
&\!\!\!\!\times\!\ldots\!\times\!\prod_{x_L}\!\sum_{y_L}\!\!\sqrt{P(y_L|x_L,\lambda(z_{0:L},s))P(y_L|x_L,\lambda(z_{0:L},s'))}
\end{align}

The final step is to use forward and reverse Gram-Schmidt procedures~\cite{dennery1996mathematics, binder2018practical} to construct the memory states, junk states and evolution operator. Notably, while the factorised memory state representation is specified in terms of an $|\mathcal{S}||\mathcal{X}|$-dimensional space, because there are only $|\mathcal{S}|$ memory states the reverse Gram-Schmidt procedure ensures that the constructed memory states inhabit only an $|\mathcal{S}|$-dimensional space. Similarly, because overlaps of junk states corresponding to different $z$ are irrelevant to the construction, the seemingly $|\mathcal{S}|(|\mathcal{X}|-1)$-dimensional junk states are actually encodable into an $|\mathcal{S}|$-dimensional space. The evolution operator $U$ then acts on this $|\mathcal{S}|^2|\mathcal{X}||\mathcal{Y}|$-dimensional joint memory-input-output-junk space.

\subsection*{G: Proof of scaling advantage}

To rigorously evaluate the memory costs of agents implementing coarse-grained strategies, we must first introduce some formal definitions. We provide definitions implicitly in terms of a single continuous parameter; the corresponding definitions for the case of coarse-graining multiple continuous parameters straightforwardly follow by nested application of the single parameter definitions. We assume the continuous parameter to be of finite domain, and without loss of generality we can take this domain to be $[0,1)$. We also explicitly consider binary coarse-grainings; the definitions and results readily generalise to arbitrary $d$-ary coarse-grainings.

\noindent {\bf Definition 3:} (Binary coarse-graining) \emph{An $n$-bit precision coarse-graining of a continuous parameter $\tau$ divides $\tau$ into $2^n$ bins of equal width $\delta \tau^{(n)}=2^{-n}$. An $n$-bit precision coarse-graining $\mathcal{P}^{(n)}$ of a strategy $\mathcal{P}$ with respect to a continuous parameter $\tau$ groups together all values of $\tau$ within each bin into a single memory state.}

A continuous parameter over the domain $[0,1)$ can be (asymptotically) represented as a binary fraction, i.e., $\tau=\sum_{k=1}^\infty \tau_k 2^{-k}$, where $\tau_k\in\{0,1\}$. Correspondingly, an $n$-bit precision coarse-graining of $\tau$, denoted by $\tau^{(n)}$, stores only the first $n$ bits of this expansion, i.e., $\tau^{(n)}=\sum_{k=1}^n\tau_k 2^{-k}$. This also provides a convenient representation for indexing the discretised bin, where the same truncated binary expansion prescribes a unique integer $\underline{\tau}^{(n)}=\sum_{k=0}^{n-1}\tau_k2^{k}$. Analogous to how the index of a causal state denotes both the label of a memory state and an equivalence class of pasts, we use $\underline{\tau}^{(n)}$ to denote both the label of the bin and the interval it spans, with the distinction clear in context. Thus, the notation $\tau\in\underline{\tau}^{(n)}$ indicates $\tau\in[\tau^{(n)},\tau^{(n)}+\delta \tau^{(n)})$. For $n>n'$, we also use the notation $\tau^{(n)}\in\underline{\tau}^{(n')}$ to indicate the set of all possible $n$-bit precision coarse-grainings of a $\tau\in\underline{\tau}^{(n')}$.

In this manner we can construct a family of coarse-grainings of a strategy at each level of precision $\{\mathcal{P}^{(n)}\}$, where $n\in\mathbb{N}$. It is implicitly assumed that such a family should converge upon the behaviour of the exact strategy in the infinite precision limit. 

When we say that an agent executes a strategy with $n$-bit precision, we mean that it has a valid encoding of an $n$-bit precision coarse-graining of the strategy. Like these coarse-grainings, we can similarly define families of agents that implement families of coarse-grainings. We denote the $n$-bit precision coarse-grained (quantum) memory states -- corresponding to the states stored by the agent executing the $n$-bit precision coarse-grained strategy -- as $\ket{\sigma_{\tau^{(n)}}^{(n)}}$ for all $\tau\in\underline{\tau}^{(n)}$. Correspondingly, we denote the overlaps of these states as \mbox{$c_{\tau^{(n)}{\tau'}^{(n)}}^{(n)}:=\braket{\sigma_{\tau^{(n)}}^{(n)}}{\sigma_{{\tau'}^{(n)}}^{(n)}}$}.

 For notational convenience in the following definitions and proof we will use the notation $P^{(n')}(\tau^{(n)})$ for \mbox{$n>n'$}, which should be interpreted as $P^{(n')}(\tau^{(n')})$ for all $\tau^{(n)}\in\underline{\tau}^{(n')}$. That is, when the argument to a coarse-grained probability is of higher-precision than the distribution, then the argument should be further coarse-grained to match the precision of the probability. An analogous interpretation should be made for the coarse-grained memory states and their overlaps, i.e., for $n>n'$, $c_{\tau^{(n)}{\tau'}^{(n)}}^{(n')}:=c_{\tau^{(n')}{\tau'}^{(n')}}^{(n')}\forall\tau^{(n)}\in\underline{\tau}^{(n')},{\tau'}^{(n)}\in{\underline{\tau}'}^{(n')}$.

With this preamble, the memory state convergence conditions can now be formally stated.

\noindent {\bf Definition 4:} (Distributional convergence) \emph{A family of coarse-grained strategies $\{\mathcal{P}^{(n)}\}$ are said to exhibit distributional convergence if for all possible input strategies $\mathcal{R}$ there exists an $n_0$ and constant $K$ such that for all $n>n_0$ the steady-states satisfy \mbox{$|P^{(n)}(\tau^{(n)})/\delta \tau^{(n)}-P^{(n-1)}(\tau^{(n)})/\delta \tau^{(n-1)}|<K\delta \tau^{(n)}\forall \tau^{(n)}$}.} 

A weaker version of this definition can be formulated, where the distributional convergence can be only with respect to a particular input strategy. If only this weaker form is satisfied, then Theorem 4 can be restated in an input strategy-dependent manner. We also note that distributional convergence implies that $P^{(n)}(\tau^{(n)})\sim\delta \tau^{(n)}$.

\noindent {\bf Definition 5:} (Memory-overlap convergence) \emph{A family of coarse-grained encoding functions $\{f^{(n)}\}$ mapping to sets of quantum memory states $\{\ket{\sigma^{(n)}_{\tau^{(n)}}}\}$ are said exhibit memory-overlap convergence if there exists an $n_0$ and constant $K$ such that for all $n>n_0$, \mbox{$|c_{\tau^{(n)}{\tau'}^{(n)}}^{(n)}-c_{\tau^{(n)}{\tau'}^{(n)}}^{(n-1)}|<K\delta \tau^{(n)}\forall \tau^{(n)},{\tau'}^{(n)}$}.} 

Armed with these definitions, we are now in a position to formally state and prove the result given in the main text regarding bounded memory costs for quantum agents executing coarse-grained strategies.

\noindent{\bf Theorem 4:} \emph{Consider a strategy $\mathcal{P}$ that has a valid encoding using memory states labelled by a finite number of continuous parameters of finite domain and a finite set of discrete parameters. A quantum adaptive agent can execute a coarse-graining of the strategy to arbitrary precision with bounded memory cost if distributional and memory-overlap convergence are satisfied.}

We first prove this for the case where the memory states are labelled by a single continuous parameter, after which we will extend to the general case.

\noindent{\bf Lemma 4:} \emph{Consider a strategy $\mathcal{P}$ that has a valid encoding using memory states labelled by a single continuous parameter of finite domain. A quantum adaptive agent can execute a coarse-graining of the strategy to arbitrary precision with bounded memory cost if distributional and memory-overlap convergence are satisfied.}

Consider such a quantum encoding at $n$-bit precision, where $n$ is sufficiently large that we above the $n_0$ required for the convergence conditions. The steady-state of the quantum agent's memory is given by $\rho^{(n)}=\sum_{\tau^{(n)}}P^{(n)}(\tau^{(n)})\ket{\sigma^{(n)}_{\tau^{(n)}}}\bra{{\sigma^{(n)}_{\tau^{(n)}}}}$. Since $\rho^{(n)}$ is finite-dimensional, the associated memory cost is finite. 

The Gram matrix~\cite{horn1990matrix} of $\rho^{(n)}$ is given by \mbox{$G^{(n)}_{\underline{\tau}^{(n)}\underline{\tau'}^{(n)}}=\sqrt{P^{(n)}(\tau^{(n)})P^{(n)}({\tau'}^{(n)})}c^{(n)}_{\tau^{(n)}{\tau'}^{(n)}}$}, and has the same spectrum (and hence von Neumann entropy) as $\rho^{(n)}$. We also define a dilated Gram matrix ${\bar{G}^{(n)}}$:
\begin{equation}
{\bar{G}^{(n)}}:=\frac{1}{2}\begin{pmatrix} 1 & 1 \\ 1 & 1\end{pmatrix}\otimes G^{(n)},
\end{equation}
such that the elements are given by \mbox{$\bar{G}^{(n)}_{\underline{\tau}^{(n+1)}\underline{\tau'}^{(n+1)}}=(1/2)\sqrt{P^{(n)}(\tau^{(n)})P^{(n)}({\tau'}^{(n)})}c^{(n)}_{\tau^{(n)}{\tau'}^{(n)}}$}.
From the properties of the tensor product, it follows that (the non-zero elements of) the spectra of $G^{(n)}$ and ${\bar{G}^{(n)}}$ are identical, and thus they have the same von Neumann entropy.

Consider also the quantum encoding at precision \mbox{$n-1$}. From the distributional and memory-overlap convergences, it follows that \mbox{$|{G^{(n)}_{\underline{\tau}^{(n)}\underline{\tau'}^{(n)}}}-\bar{G}^{(n-1)}_{\underline{\tau}^{(n)}\underline{\tau'}^{(n)}}|<K{\delta \tau^{(n)}}^2$} $\forall \underline{\tau}^{(n)},\underline{\tau'}^{(n)}$ for some constant $K$. We define this matrix of differences $\Delta^{(n)}:=G^{(n)}-{\bar{G}^{(n-1)}}$; its elements scale as $O(2^{-2n})$.

The Schatten $p$-norms of a matrix $A$ are defined $\|A\|_p:=\Tr(|A|^p)^{\frac{1}{p}}$ for $p\in[1,\infty)$~\cite{watrous2018theory}. They satisfy H\"{o}lder's inequality, whereby $\|AB\|_1\leq\|A\|_p\|B\|_q$ for \mbox{$1/p+1/q=1$}. Two special cases of relevance here are $p=1$, also referred to as the trace norm, and $p=2$, which is equivalent to the Frobenius norm \mbox{$\|A\|_F:=\sqrt{\sum_{jk}|A_{jk}|^2}=\|A\|_2$}. Noting that $\Delta^{(n)}$ has $2^{n}\times2^{n}$ elements, we have that $\|\Delta^{(n)}\|_2\sim 2^{-n}$. Then, by applying H\"{o}lder's inequality with $p=2$, \mbox{$A=\Delta^{(n)}$}, and $B$ the identity matrix over the space occupied by $\Delta^{(n)}$, we have that $\|\Delta^{(n)}\|_1\sim 2^{-n/2}$.

The Fannes-Audenaert inequality~\cite{watrous2018theory} relates the difference in von Neumann entropies of two operators with the trace norm of their difference. For two operators $\rho_A$ and $\rho_B$ of dimension $d$, it states that 
\begin{align}
|S_{\mathrm{vN}}[\rho_A]-S_{\mathrm{vN}}[\rho_B]|\leq\frac{1}{2}&\log_2(d-1)\|\rho_A-\rho_B\|_1\nonumber\\&+h\left(\frac{1}{2}\|\rho_A-\rho_B\|_1\right),
\end{align}
where $h(x):=-x\log_2(x)-(1-x)\log_2(1-x)$. Setting $\rho_A=G^{(n)}$ and $\rho_B={\bar{G}^{(n-1)}}$, together with the above we arrive at
\begin{equation}
|S_{\mathrm{vN}}[\rho^{(n)}]-S_{\mathrm{vN}}[\rho^{(n-1)}]|\leq \sim n2^{-\frac{n}{2}}.
\end{equation}
Thus, beyond a sufficiently high precision, the increase in the quantum agent's memory cost for each extra degree of precision is exponentially-decreasing. Correspondingly, the memory cost will eventually converge when the precision is increased an arbitrary number of times, leading to a bounded memory cost at any level of precision.

When we have an additional set of discrete parameters $m\in\mathcal{M}$ labelling the memory states, such that the pair $(t,m)\in(\tau,\mathcal{M})$ uniquely specifies the memory state, we effectively have a finite number of sectors for the memory state space, with each sector corresponding to a different $m$. The state convergence conditions readily generalise to this regime, by imposing the conditions on each sector individually. Then, by applying the above arguments in the proof of Lemma 4 to each sector, we see that the total contribution to the memory cost from the memory states in each sector is bounded. Since there are a finite number of sectors, the total memory cost is thus bounded.

When there are multiple continuous parameters, the conditions on convergence must apply to all such parameters. Beginning from a sufficiently fine discretisation of all continuous parameters, we can apply the arguments above to each continuous parameter in turn, to deduce that the memory cost remains bounded at arbitrary precision in all continuous parameters. This completes the proof of Theorem 4.

Finally, we remark that while we have assumed finite, discrete stimulus and action alphabets in the above, the definitions and proofs readily extend to the case where these also are continuous parameters of finite domain.

\subsection*{H: Details for resettable stochastic clocks}

A renewal process~\cite{smith1958renewal} is described by a series of identical events, where the time interval between consecutive events is drawn randomly from a distribution $\phi(t)$; here we focus on the case where this is discretised into timesteps of size $\delta t$. A resettable renewal process can accept input stimuli that trigger a `reset' of the system to its post-event state, in effect triggering a phantom event and restarting the timer to the next event. We can describe the input stimulus by a two symbol alphabet: 0 (\texttt{continue}) and 1 (\texttt{reset}). Similarly, the output action alphabet can be described by two symbols: 0 for non-events and 1 for events. $\Phi(t):=\int_t^\infty\phi(t')dt'$ (and discrete analogue thereof) represents the so-called survival probability of the process. Such resettable renewal processes correspond to the strategy of resettable stochastic clocks.

It is clear that since the agent will always behave the same on stimulus 1, the groupings of pasts into causal states depends only on their response to stimulus 0. This recovers the vanilla renewal process case, and we obtain the same causal states as in such settings~\cite{marzen2015informational, marzen2017informational, elliott2018superior, elliott2020extreme}: outside of specific forms of $\phi(t)$ -- that we shall ignore here, noting that the following analysis can straightforwardly be generalised to encompass them -- the causal states $s_n$ of a renewal process describe the number of timesteps $n$ since the last event (in our case, this also includes the phantom events from resets). 

The steady-state distribution of the causal states can be readily calculated for any resettable renewal process where the input stimuli are themselves driven by an input renewal process that resets upon events from either process. Label the event distribution and survival probability of the input process as $\phi_I(t)$ and $\Phi_I(t)$ respectively, and similarly $\phi_O(t)$ and $\Phi_O(t)$ for the strategy renewal process. For a pure renewal process without resettability, the steady-state distribution is given by $\mu\Phi(t)$, where the normalisation $\mu:=\left(\int_0^\infty t\phi(t)dt\right)^{-1}=\left(\int_0^\infty\Phi(t)dt\right)^{-1}$ (replace integrals with sums for the discrete-time case) is called the mean-firing rate, and represents the average number of events per unit time/timestep~\cite{marzen2015informational, marzen2017informational, elliott2018superior}. Since both processes are reset by events on the strategy process, we can view the pure output action process without reference to the input as a renewal process in its own right, with an effective event distribution being a function of both stimulus and action event distributions. The effective survival probability is the product of the survival probabilities, as the pure output process will only survive up to a given time provided that neither the underlying renewal process or the input renewal process have fired. Thus, the steady-state probabilities will be proportional to $\Phi_I(t)\Phi_O(t)$, and normalised by their sum/integral, which yields the effective mean firing rate. With these probabilities, the (input-dependent) minimal classical memory cost can be straightforwardly calculated.

To determine the corresponding memory measure for our quantum agent we must also calculate the memory state overlaps. Using Eq.~\eqref{eqoverlap}, and noting that all causal states behave identically on input 1 we obtain
\begin{equation}
\label{eqquantumnewoverlap}
c_{tt'}\!=\!\!\sqrt{\!\frac{\Phi_O(t\!+\!\delta t)\Phi_O(t'\!\!+\!\delta t)}{\Phi_O(t)\Phi_O(t')}}c_{(t+\delta t)(t'+\delta t)}\!+\!\sqrt{\!\frac{\bar{\phi_O}(t)\bar{\phi_O}(t')}{\Phi_O(t)\Phi_O(t')}}.
\end{equation}
where \mbox{$\bar{\phi}_O(t):=\int_{t}^{t+\delta t}\phi_O(t)dt=\Phi_O(t)-\Phi_O(t+\delta t)$}. From these iterative equations we obtain
\begin{equation}
\label{eqquantumnewoverlap2}
c_{tt'}=\sum_{j=0}^{\infty}\sqrt{\frac{\bar{\phi_O}(t+j\delta t)\bar{\phi_O}(t'+j\delta t)}{\Phi_O(t)\Phi_O(t')}}.
\end{equation}
These overlaps saturate the fidelity bound Eq.~\eqref{eqbound}. Together with the steady-state probabilities, we can calculate the input-dependent memory cost of our agent.

We also compare with the prior proof of principle quantum agent~\cite{thompson2017using}. To determine the overlaps of its memory states $\{\ket{S_s}:=\bigotimes_{x}\ket{S^{x}_s}\}$ we recast this agent in terms of our general form Eq.~\eqref{eqgeneral}:
\begin{align}
U_{q_1}\ket{S_s}\ket{x}\ket{0}\ket{0}=\sum_y&\sqrt{P(y|x,s)}\ket{S_{\lambda(z,s)}}\ket{x}\ket{y}\nonumber\\&\bigotimes_{x'\neq x}\ket{S^{x'}_s}\ket{\lambda(z,s)}.
\end{align}
We note that this formulation in terms of unitary evolution differs from the original presentation, though yields identical memory states. Expressed in this way it is clear where the deficiency of this agent relative to ours lies -- in announcing the next causal state in the junk. The corresponding overlaps between the memory states for resettable stochastic clocks are given by
\begin{equation}
\label{eqquantumoldoverlap}
c_{tt'}\!=\!\!\sqrt{\!\frac{\Phi_O(t\!+\!\delta t)\Phi_O(t'\!\!+\!\delta t)}{\Phi_O(t)\Phi_O(t')}}\delta_{(t+\delta t)(t'+\delta t)}\!+\!\sqrt{\!\frac{\bar{\phi_O}(t)\bar{\phi_O}(t')}{\Phi_O(t)\Phi_O(t')}}.
\end{equation}
With these overlaps, we can calculate the memory requirement of the agent. It can be seen that the overlaps rely not only on overlap in the output action statistics, but also that the immediately subsequent causal state into which the system transitions is the same. In contrast, our agent relies only on the overlap of output action statistics (over arbitrarily long horizons), which due to asymptotic synchronisation to a causal state over sufficiently long pasts requires that the transition into the same causal state is mandated only for arbitrarily far into the future. It is for this reason that we label the quantum agents with subscripts 1 and $\infty$, and it becomes clear why our new agent drastically outperforms the prior agent for processes with long historical dependence.

Indeed, the example presented in the main text is not an isolated case of the scaling advantage for a particular resettable stochastic clock. Theorem 4 provides us with a sufficiency condition against which we can verify that typical resettable stochastic clocks with a smooth distribution $\phi(t)$ requires only a bounded amount of memory to execute when driven by a smooth renewal process.

\noindent{\bf Corollary 1:} \emph{Consider a resettable stochastic clock with distribution $\phi_O(t)$ that is either of finite domain or takes the form of a Poisson process at long times. Suppose that it is driven by a renewal process with distribution $\phi_I(t)$ that resets upon clock ticks. If $\Phi_O(t)$ and $\Phi_I(t)$ are infinitely differentiable, then a quantum agent encoded using Algorithm 1 can execute the strategy to arbitrary precision with only a bounded memory cost.}

Let us begin by considering the case where $\phi_O(t)$ is of finite domain. Since $\phi=-d\Phi/dt$, we also have that the distributions $\phi_O(t)$ and $\phi_I(t)$ are infinitely differentiable, and so can be approximated over small distances by a Taylor expansion (i.e., $\phi(t+\delta t)\approx\phi(t)+\partial_t\phi(t)\delta t$). Then, the coarse-graining $P^{(n)}(1|0,t^{(n)})=\bar{\phi}^{(n)}_O(t^{(n)})=\int_{\underline{t}^{(n)}}\phi_O(t)dt$ can readily be verified to exponentially converge towards the exact strategy with increasing precision. We also have that $\Phi^{(n)}(t^{(n)})=\Phi(t^{(n)})$ $\forall n\in\mathbb{N}$, since 
\begin{align}
\Phi^{(n)}(t^{(n)}):=&\sum_{j=0}^\infty \bar{\phi}^{(n)}(t^{(n)}+j\delta t^{(n)})\nonumber\\
=&\sum_{j=0}^\infty\int_{\underline{t^{(n)}+j\delta t^{(n)}}}\phi(t')dt'\nonumber\\
=&\int_{t^{(n)}}^\infty\phi(t')dt'=\Phi(t^{(n)}).
\end{align}

Consider now the overlaps $c_{tt'}^{(n)}$ as prescribed in Eq.~\eqref{eqquantumnewoverlap2}. We can make an expansion $\bar{\phi}^{(n)}(t^{(n)}+\delta t^{n})\approx \bar{\phi}^{(n)}(t^{(n)})+\partial_t\bar{\phi}^{(n)}(t^{(n)})\delta t^{(n)}$, where the correction is $O({\delta t^{(n)}}^2)$ due to $\bar{\phi}^{(n)}(t^{(n)})$ scaling as $O(\delta t^{(n)})$. Thus, for sufficiently large $n$, $\bar{\phi}^{(n)}(t^{(n)})\approx\bar{\phi}^{(n)}(t^{(n)}+\delta t^{(n)})\approx\bar{\phi}^{(n-1)}(t^{(n)})/2+O({\delta t^{(n)}}^2)$. Since there at most $2^n$ non-zero terms in the sum constituting $c^{(n)}_{tt'}$, it then follows that at sufficiently large $n$, \mbox{$|c^{(n)}_{tt'}-c^{(n-1)}_{tt'}|\leq\sim\delta t^{(n)}\forall t,t'$}; memory-overlap convergence is satisfied.

The steady-state probabilities are given by $P^{(n)}(t^{(n)})=\mu_{IO}^{(n)}\Phi_I(t^{(n)})\Phi_O(t^{(n)})$, where ${\mu_{IO}^{(n)}}^{-1}:=\sum_{j=0}^\infty\Phi_I(j\delta t^{(n)})\Phi_O(j\delta t^{(n)})$. For sufficiently large $n$, it follows that ${\mu_{IO}^{(n)}}^{-1}\approx 2{\mu_{IO}^{(n-1)}}^{-1}-1/2$, and thus ${\mu_{IO}^{(n)}}\sim\delta t^{(n)}$. It can then readily be verified that distributional convergence is satisfied. Hence, the convergence conditions of Theorem 4 are satisfied, and it therefore follows that the memory cost remains bounded, irrespective of the precision.

More generally, when $\phi_O(t)$ takes the form of a Poisson process at long times (say, for $t>\tau_0$), we can partition the memory states in two, according to whether $t$ is below or above $\tau_0$. This binary classification defines two sectors of memory states. For the former, we can apply the above arguments to show that in this sector the convergence conditions are satisfied. Meanwhile, we can apply known properties of the causal states of renewal processes~\cite{marzen2015informational} to see that all memory states in the latter sector belong to the same causal state, and hence Algorithm 1 maps them all to identical states. Thus, the convergence conditions are also satisfied in this sector, and hence Theorem 4 can again be applied.

\acknowledgments
This work was funded by the Imperial College Borland Fellowship in Mathematics, the Lee Kuan Yew Endowment Fund (Postdoctoral Fellowship), grants FQXi-RFP-1809 and FQXi-RFP-IPW-1903 from the Foundational Questions Institute and Fetzer Franklin Fund (a donor advised fund of Silicon Valley Community Foundation), Singapore Ministry of Education through Tier 1 grant no.~RG190/17, and the Singapore National Research Foundation through Fellowship no.~NRF-NRFF2016-02 and NRF-ANR grant no.~NRF2017-NRF-ANR004 VanQuTe. T.J.E.~thanks the Centre for Quantum Technologies for their hospitality.

\bibliography{ref}

\begin{thebibliography}{73}%
\makeatletter
\providecommand \@ifxundefined [1]{%
 \@ifx{#1\undefined}
}%
\providecommand \@ifnum [1]{%
 \ifnum #1\expandafter \@firstoftwo
 \else \expandafter \@secondoftwo
 \fi
}%
\providecommand \@ifx [1]{%
 \ifx #1\expandafter \@firstoftwo
 \else \expandafter \@secondoftwo
 \fi
}%
\providecommand \natexlab [1]{#1}%
\providecommand \enquote  [1]{``#1''}%
\providecommand \bibnamefont  [1]{#1}%
\providecommand \bibfnamefont [1]{#1}%
\providecommand \citenamefont [1]{#1}%
\providecommand \href@noop [0]{\@secondoftwo}%
\providecommand \href [0]{\begingroup \@sanitize@url \@href}%
\providecommand \@href[1]{\@@startlink{#1}\@@href}%
\providecommand \@@href[1]{\endgroup#1\@@endlink}%
\providecommand \@sanitize@url [0]{\catcode `\\12\catcode `\$12\catcode
  `\&12\catcode `\#12\catcode `\^12\catcode `\_12\catcode `\%12\relax}%
\providecommand \@@startlink[1]{}%
\providecommand \@@endlink[0]{}%
\providecommand \url  [0]{\begingroup\@sanitize@url \@url }%
\providecommand \@url [1]{\endgroup\@href {#1}{\urlprefix }}%
\providecommand \urlprefix  [0]{URL }%
\providecommand \Eprint [0]{\href }%
\providecommand \doibase [0]{https://doi.org/}%
\providecommand \selectlanguage [0]{\@gobble}%
\providecommand \bibinfo  [0]{\@secondoftwo}%
\providecommand \bibfield  [0]{\@secondoftwo}%
\providecommand \translation [1]{[#1]}%
\providecommand \BibitemOpen [0]{}%
\providecommand \bibitemStop [0]{}%
\providecommand \bibitemNoStop [0]{.\EOS\space}%
\providecommand \EOS [0]{\spacefactor3000\relax}%
\providecommand \BibitemShut  [1]{\csname bibitem#1\endcsname}%
\let\auto@bib@innerbib\@empty
\bibitem [{\citenamefont {Bonabeau}(2002)}]{bonabeau2002agent}%
  \BibitemOpen
  \bibfield  {author} {\bibinfo {author} {\bibfnamefont {E.}~\bibnamefont
  {Bonabeau}},\ }\bibfield  {title} {\bibinfo {title} {Agent-based modeling:
  Methods and techniques for simulating human systems},\ }\href@noop {}
  {\bibfield  {journal} {\bibinfo  {journal} {Proceedings of the National
  Academy of Sciences}\ }\textbf {\bibinfo {volume} {99}},\ \bibinfo {pages}
  {7280} (\bibinfo {year} {2002})}\BibitemShut {NoStop}%
\bibitem [{\citenamefont {Macy}\ and\ \citenamefont
  {Willer}(2002)}]{macy2002factors}%
  \BibitemOpen
  \bibfield  {author} {\bibinfo {author} {\bibfnamefont {M.~W.}\ \bibnamefont
  {Macy}}\ and\ \bibinfo {author} {\bibfnamefont {R.}~\bibnamefont {Willer}},\
  }\bibfield  {title} {\bibinfo {title} {From factors to actors: Computational
  sociology and agent-based modeling},\ }\href@noop {} {\bibfield  {journal}
  {\bibinfo  {journal} {Annual Review of Sociology}\ }\textbf {\bibinfo
  {volume} {28}},\ \bibinfo {pages} {143} (\bibinfo {year} {2002})}\BibitemShut
  {NoStop}%
\bibitem [{\citenamefont {Macal}\ and\ \citenamefont
  {North}(2005)}]{macal2005tutorial}%
  \BibitemOpen
  \bibfield  {author} {\bibinfo {author} {\bibfnamefont {C.~M.}\ \bibnamefont
  {Macal}}\ and\ \bibinfo {author} {\bibfnamefont {M.~J.}\ \bibnamefont
  {North}},\ }\bibfield  {title} {\bibinfo {title} {Tutorial on agent-based
  modeling and simulation},\ }in\ \href@noop {} {\emph {\bibinfo {booktitle}
  {Proceedings of the Winter Simulation Conference, 2005.}}}\ (\bibinfo
  {organization} {IEEE},\ \bibinfo {year} {2005})\ p.~\bibinfo {pages}
  {14}\BibitemShut {NoStop}%
\bibitem [{\citenamefont {Wooldridge}(2009)}]{wooldridge2009introduction}%
  \BibitemOpen
  \bibfield  {author} {\bibinfo {author} {\bibfnamefont {M.}~\bibnamefont
  {Wooldridge}},\ }\href@noop {} {\emph {\bibinfo {title} {An introduction to
  multiagent systems}}}\ (\bibinfo  {publisher} {John Wiley \& Sons},\ \bibinfo
  {year} {2009})\BibitemShut {NoStop}%
\bibitem [{\citenamefont {An}(2012)}]{an2012modeling}%
  \BibitemOpen
  \bibfield  {author} {\bibinfo {author} {\bibfnamefont {L.}~\bibnamefont
  {An}},\ }\bibfield  {title} {\bibinfo {title} {Modeling human decisions in
  coupled human and natural systems: Review of agent-based models},\
  }\href@noop {} {\bibfield  {journal} {\bibinfo  {journal} {Ecological
  Modelling}\ }\textbf {\bibinfo {volume} {229}},\ \bibinfo {pages} {25}
  (\bibinfo {year} {2012})}\BibitemShut {NoStop}%
\bibitem [{\citenamefont {Farmer}\ and\ \citenamefont
  {Foley}(2009)}]{farmer2009economy}%
  \BibitemOpen
  \bibfield  {author} {\bibinfo {author} {\bibfnamefont {J.~D.}\ \bibnamefont
  {Farmer}}\ and\ \bibinfo {author} {\bibfnamefont {D.}~\bibnamefont {Foley}},\
  }\bibfield  {title} {\bibinfo {title} {The economy needs agent-based
  modelling},\ }\href@noop {} {\bibfield  {journal} {\bibinfo  {journal}
  {Nature}\ }\textbf {\bibinfo {volume} {460}},\ \bibinfo {pages} {685}
  (\bibinfo {year} {2009})}\BibitemShut {NoStop}%
\bibitem [{\citenamefont {Thurner}\ \emph {et~al.}(2012)\citenamefont
  {Thurner}, \citenamefont {Farmer},\ and\ \citenamefont
  {Geanakoplos}}]{thurner2012leverage}%
  \BibitemOpen
  \bibfield  {author} {\bibinfo {author} {\bibfnamefont {S.}~\bibnamefont
  {Thurner}}, \bibinfo {author} {\bibfnamefont {J.~D.}\ \bibnamefont
  {Farmer}},\ and\ \bibinfo {author} {\bibfnamefont {J.}~\bibnamefont
  {Geanakoplos}},\ }\bibfield  {title} {\bibinfo {title} {Leverage causes fat
  tails and clustered volatility},\ }\href@noop {} {\bibfield  {journal}
  {\bibinfo  {journal} {Quantitative Finance}\ }\textbf {\bibinfo {volume}
  {12}},\ \bibinfo {pages} {695} (\bibinfo {year} {2012})}\BibitemShut
  {NoStop}%
\bibitem [{\citenamefont {Lardon}\ \emph {et~al.}(2011)\citenamefont {Lardon},
  \citenamefont {Merkey}, \citenamefont {Martins}, \citenamefont {D{\"o}tsch},
  \citenamefont {Picioreanu}, \citenamefont {Kreft},\ and\ \citenamefont
  {Smets}}]{lardon2011idynomics}%
  \BibitemOpen
  \bibfield  {author} {\bibinfo {author} {\bibfnamefont {L.~A.}\ \bibnamefont
  {Lardon}}, \bibinfo {author} {\bibfnamefont {B.~V.}\ \bibnamefont {Merkey}},
  \bibinfo {author} {\bibfnamefont {S.}~\bibnamefont {Martins}}, \bibinfo
  {author} {\bibfnamefont {A.}~\bibnamefont {D{\"o}tsch}}, \bibinfo {author}
  {\bibfnamefont {C.}~\bibnamefont {Picioreanu}}, \bibinfo {author}
  {\bibfnamefont {J.-U.}\ \bibnamefont {Kreft}},\ and\ \bibinfo {author}
  {\bibfnamefont {B.~F.}\ \bibnamefont {Smets}},\ }\bibfield  {title} {\bibinfo
  {title} {{iDynoMiCS}: Next-generation individual-based modelling of
  biofilms},\ }\href@noop {} {\bibfield  {journal} {\bibinfo  {journal}
  {Environmental Microbiology}\ }\textbf {\bibinfo {volume} {13}},\ \bibinfo
  {pages} {2416} (\bibinfo {year} {2011})}\BibitemShut {NoStop}%
\bibitem [{\citenamefont {Mei}\ \emph {et~al.}(2010)\citenamefont {Mei},
  \citenamefont {Sloot}, \citenamefont {Quax}, \citenamefont {Zhu},\ and\
  \citenamefont {Wang}}]{mei2010complex}%
  \BibitemOpen
  \bibfield  {author} {\bibinfo {author} {\bibfnamefont {S.}~\bibnamefont
  {Mei}}, \bibinfo {author} {\bibfnamefont {P.~M.~A.}\ \bibnamefont {Sloot}},
  \bibinfo {author} {\bibfnamefont {R.}~\bibnamefont {Quax}}, \bibinfo {author}
  {\bibfnamefont {Y.}~\bibnamefont {Zhu}},\ and\ \bibinfo {author}
  {\bibfnamefont {W.}~\bibnamefont {Wang}},\ }\bibfield  {title} {\bibinfo
  {title} {Complex agent networks explaining the hiv epidemic among homosexual
  men in amsterdam},\ }\href@noop {} {\bibfield  {journal} {\bibinfo  {journal}
  {Mathematics and Computers in Simulation}\ }\textbf {\bibinfo {volume}
  {80}},\ \bibinfo {pages} {1018} (\bibinfo {year} {2010})}\BibitemShut
  {NoStop}%
\bibitem [{\citenamefont {Cassandra}\ \emph {et~al.}(1994)\citenamefont
  {Cassandra}, \citenamefont {Kaelbling},\ and\ \citenamefont
  {Littman}}]{cassandra1994acting}%
  \BibitemOpen
  \bibfield  {author} {\bibinfo {author} {\bibfnamefont {A.~R.}\ \bibnamefont
  {Cassandra}}, \bibinfo {author} {\bibfnamefont {L.~P.}\ \bibnamefont
  {Kaelbling}},\ and\ \bibinfo {author} {\bibfnamefont {M.~L.}\ \bibnamefont
  {Littman}},\ }\bibfield  {title} {\bibinfo {title} {Acting optimally in
  partially observable stochastic domains},\ }in\ \href@noop {} {\emph
  {\bibinfo {booktitle} {Proceedings of the Twelfth AAAI National Conference on
  Artificial Intelligence}}}\ (\bibinfo {organization} {AAAI Press},\ \bibinfo
  {year} {1994})\ pp.\ \bibinfo {pages} {1023--1028}\BibitemShut {NoStop}%
\bibitem [{\citenamefont {Thompson}\ \emph {et~al.}(2017)\citenamefont
  {Thompson}, \citenamefont {Garner}, \citenamefont {Vedral},\ and\
  \citenamefont {Gu}}]{thompson2017using}%
  \BibitemOpen
  \bibfield  {author} {\bibinfo {author} {\bibfnamefont {J.}~\bibnamefont
  {Thompson}}, \bibinfo {author} {\bibfnamefont {A.~J.~P.}\ \bibnamefont
  {Garner}}, \bibinfo {author} {\bibfnamefont {V.}~\bibnamefont {Vedral}},\
  and\ \bibinfo {author} {\bibfnamefont {M.}~\bibnamefont {Gu}},\ }\bibfield
  {title} {\bibinfo {title} {Using quantum theory to simplify input-output
  processes},\ }\href@noop {} {\bibfield  {journal} {\bibinfo  {journal} {npj
  Quantum Information}\ }\textbf {\bibinfo {volume} {3}},\ \bibinfo {pages} {6}
  (\bibinfo {year} {2017})}\BibitemShut {NoStop}%
\bibitem [{\citenamefont {Gu}\ \emph {et~al.}(2012)\citenamefont {Gu},
  \citenamefont {Wiesner}, \citenamefont {Rieper},\ and\ \citenamefont
  {Vedral}}]{gu2012quantum}%
  \BibitemOpen
  \bibfield  {author} {\bibinfo {author} {\bibfnamefont {M.}~\bibnamefont
  {Gu}}, \bibinfo {author} {\bibfnamefont {K.}~\bibnamefont {Wiesner}},
  \bibinfo {author} {\bibfnamefont {E.}~\bibnamefont {Rieper}},\ and\ \bibinfo
  {author} {\bibfnamefont {V.}~\bibnamefont {Vedral}},\ }\bibfield  {title}
  {\bibinfo {title} {Quantum mechanics can reduce the complexity of classical
  models},\ }\href@noop {} {\bibfield  {journal} {\bibinfo  {journal} {Nature
  Communications}\ }\textbf {\bibinfo {volume} {3}},\ \bibinfo {pages} {762}
  (\bibinfo {year} {2012})}\BibitemShut {NoStop}%
\bibitem [{\citenamefont {Mahoney}\ \emph {et~al.}(2016)\citenamefont
  {Mahoney}, \citenamefont {Aghamohammadi},\ and\ \citenamefont
  {Crutchfield}}]{mahoney2016occam}%
  \BibitemOpen
  \bibfield  {author} {\bibinfo {author} {\bibfnamefont {J.~R.}\ \bibnamefont
  {Mahoney}}, \bibinfo {author} {\bibfnamefont {C.}~\bibnamefont
  {Aghamohammadi}},\ and\ \bibinfo {author} {\bibfnamefont {J.~P.}\
  \bibnamefont {Crutchfield}},\ }\bibfield  {title} {\bibinfo {title} {Occam's
  quantum strop: {S}ynchronizing and compressing classical cryptic processes
  via a quantum channel},\ }\href@noop {} {\bibfield  {journal} {\bibinfo
  {journal} {Scientific Reports}\ }\textbf {\bibinfo {volume} {6}},\ \bibinfo
  {pages} {20495} (\bibinfo {year} {2016})}\BibitemShut {NoStop}%
\bibitem [{\citenamefont {Aghamohammadi}\ \emph {et~al.}(2018)\citenamefont
  {Aghamohammadi}, \citenamefont {Loomis}, \citenamefont {Mahoney},\ and\
  \citenamefont {Crutchfield}}]{aghamohammadi2018extreme}%
  \BibitemOpen
  \bibfield  {author} {\bibinfo {author} {\bibfnamefont {C.}~\bibnamefont
  {Aghamohammadi}}, \bibinfo {author} {\bibfnamefont {S.~P.}\ \bibnamefont
  {Loomis}}, \bibinfo {author} {\bibfnamefont {J.~R.}\ \bibnamefont
  {Mahoney}},\ and\ \bibinfo {author} {\bibfnamefont {J.~P.}\ \bibnamefont
  {Crutchfield}},\ }\bibfield  {title} {\bibinfo {title} {Extreme quantum
  memory advantage for rare-event sampling},\ }\href@noop {} {\bibfield
  {journal} {\bibinfo  {journal} {Physical Review X}\ }\textbf {\bibinfo
  {volume} {8}},\ \bibinfo {pages} {011025} (\bibinfo {year}
  {2018})}\BibitemShut {NoStop}%
\bibitem [{\citenamefont {Elliott}\ and\ \citenamefont
  {Gu}(2018)}]{elliott2018superior}%
  \BibitemOpen
  \bibfield  {author} {\bibinfo {author} {\bibfnamefont {T.~J.}\ \bibnamefont
  {Elliott}}\ and\ \bibinfo {author} {\bibfnamefont {M.}~\bibnamefont {Gu}},\
  }\bibfield  {title} {\bibinfo {title} {Superior memory efficiency of quantum
  devices for the simulation of continuous-time stochastic processes},\
  }\href@noop {} {\bibfield  {journal} {\bibinfo  {journal} {npj Quantum
  Information}\ }\textbf {\bibinfo {volume} {4}},\ \bibinfo {pages} {18}
  (\bibinfo {year} {2018})}\BibitemShut {NoStop}%
\bibitem [{\citenamefont {Binder}\ \emph {et~al.}(2018)\citenamefont {Binder},
  \citenamefont {Thompson},\ and\ \citenamefont {Gu}}]{binder2018practical}%
  \BibitemOpen
  \bibfield  {author} {\bibinfo {author} {\bibfnamefont {F.~C.}\ \bibnamefont
  {Binder}}, \bibinfo {author} {\bibfnamefont {J.}~\bibnamefont {Thompson}},\
  and\ \bibinfo {author} {\bibfnamefont {M.}~\bibnamefont {Gu}},\ }\bibfield
  {title} {\bibinfo {title} {Practical unitary simulator for non-{M}arkovian
  complex processes},\ }\href@noop {} {\bibfield  {journal} {\bibinfo
  {journal} {Physical Review Letters}\ }\textbf {\bibinfo {volume} {120}},\
  \bibinfo {pages} {240502} (\bibinfo {year} {2018})}\BibitemShut {NoStop}%
\bibitem [{\citenamefont {Elliott}\ \emph {et~al.}(2019)\citenamefont
  {Elliott}, \citenamefont {Garner},\ and\ \citenamefont
  {Gu}}]{elliott2019memory}%
  \BibitemOpen
  \bibfield  {author} {\bibinfo {author} {\bibfnamefont {T.~J.}\ \bibnamefont
  {Elliott}}, \bibinfo {author} {\bibfnamefont {A.~J.~P.}\ \bibnamefont
  {Garner}},\ and\ \bibinfo {author} {\bibfnamefont {M.}~\bibnamefont {Gu}},\
  }\bibfield  {title} {\bibinfo {title} {{Memory-efficient tracking of complex
  temporal and symbolic dynamics with quantum simulators}},\ }\href@noop {}
  {\bibfield  {journal} {\bibinfo  {journal} {New Journal of Physics}\ }\textbf
  {\bibinfo {volume} {21}},\ \bibinfo {pages} {013021} (\bibinfo {year}
  {2019})}\BibitemShut {NoStop}%
\bibitem [{\citenamefont {Thompson}\ \emph {et~al.}(2018)\citenamefont
  {Thompson}, \citenamefont {Garner}, \citenamefont {Mahoney}, \citenamefont
  {Crutchfield}, \citenamefont {Vedral},\ and\ \citenamefont
  {Gu}}]{thompson2018causal}%
  \BibitemOpen
  \bibfield  {author} {\bibinfo {author} {\bibfnamefont {J.}~\bibnamefont
  {Thompson}}, \bibinfo {author} {\bibfnamefont {A.~J.~P.}\ \bibnamefont
  {Garner}}, \bibinfo {author} {\bibfnamefont {J.~R.}\ \bibnamefont {Mahoney}},
  \bibinfo {author} {\bibfnamefont {J.~P.}\ \bibnamefont {Crutchfield}},
  \bibinfo {author} {\bibfnamefont {V.}~\bibnamefont {Vedral}},\ and\ \bibinfo
  {author} {\bibfnamefont {M.}~\bibnamefont {Gu}},\ }\bibfield  {title}
  {\bibinfo {title} {Causal asymmetry in a quantum world},\ }\href@noop {}
  {\bibfield  {journal} {\bibinfo  {journal} {Physical Review X}\ }\textbf
  {\bibinfo {volume} {8}},\ \bibinfo {pages} {031013} (\bibinfo {year}
  {2018})}\BibitemShut {NoStop}%
\bibitem [{\citenamefont {Elliott}\ \emph {et~al.}(2020)\citenamefont
  {Elliott}, \citenamefont {Yang}, \citenamefont {Binder}, \citenamefont
  {Garner}, \citenamefont {Thompson},\ and\ \citenamefont
  {Gu}}]{elliott2020extreme}%
  \BibitemOpen
  \bibfield  {author} {\bibinfo {author} {\bibfnamefont {T.~J.}\ \bibnamefont
  {Elliott}}, \bibinfo {author} {\bibfnamefont {C.}~\bibnamefont {Yang}},
  \bibinfo {author} {\bibfnamefont {F.~C.}\ \bibnamefont {Binder}}, \bibinfo
  {author} {\bibfnamefont {A.~J.~P.}\ \bibnamefont {Garner}}, \bibinfo {author}
  {\bibfnamefont {J.}~\bibnamefont {Thompson}},\ and\ \bibinfo {author}
  {\bibfnamefont {M.}~\bibnamefont {Gu}},\ }\bibfield  {title} {\bibinfo
  {title} {Extreme dimensionality reduction with quantum modeling},\
  }\href@noop {} {\bibfield  {journal} {\bibinfo  {journal} {Physical Review
  Letters}\ }\textbf {\bibinfo {volume} {125}},\ \bibinfo {pages} {260501}
  (\bibinfo {year} {2020})}\BibitemShut {NoStop}%
\bibitem [{\citenamefont {Liu}\ \emph {et~al.}(2019)\citenamefont {Liu},
  \citenamefont {Elliott}, \citenamefont {Binder}, \citenamefont {Di~Franco},\
  and\ \citenamefont {Gu}}]{liu2019optimal}%
  \BibitemOpen
  \bibfield  {author} {\bibinfo {author} {\bibfnamefont {Q.}~\bibnamefont
  {Liu}}, \bibinfo {author} {\bibfnamefont {T.~J.}\ \bibnamefont {Elliott}},
  \bibinfo {author} {\bibfnamefont {F.~C.}\ \bibnamefont {Binder}}, \bibinfo
  {author} {\bibfnamefont {C.}~\bibnamefont {Di~Franco}},\ and\ \bibinfo
  {author} {\bibfnamefont {M.}~\bibnamefont {Gu}},\ }\bibfield  {title}
  {\bibinfo {title} {Optimal stochastic modeling with unitary quantum
  dynamics},\ }\href@noop {} {\bibfield  {journal} {\bibinfo  {journal}
  {Physical Review A}\ }\textbf {\bibinfo {volume} {99}},\ \bibinfo {pages}
  {062110} (\bibinfo {year} {2019})}\BibitemShut {NoStop}%
\bibitem [{\citenamefont {Barnett}\ and\ \citenamefont
  {Crutchfield}(2015)}]{barnett2015computational}%
  \BibitemOpen
  \bibfield  {author} {\bibinfo {author} {\bibfnamefont {N.}~\bibnamefont
  {Barnett}}\ and\ \bibinfo {author} {\bibfnamefont {J.~P.}\ \bibnamefont
  {Crutchfield}},\ }\bibfield  {title} {\bibinfo {title} {Computational
  mechanics of input--output processes: Structured transformations and the
  $\varepsilon$-transducer},\ }\href@noop {} {\bibfield  {journal} {\bibinfo
  {journal} {Journal of Statistical Physics}\ }\textbf {\bibinfo {volume}
  {161}},\ \bibinfo {pages} {404} (\bibinfo {year} {2015})}\BibitemShut
  {NoStop}%
\bibitem [{\citenamefont {Dong}\ \emph {et~al.}(2008)\citenamefont {Dong},
  \citenamefont {Chen}, \citenamefont {Li},\ and\ \citenamefont
  {Tarn}}]{dong2008quantum}%
  \BibitemOpen
  \bibfield  {author} {\bibinfo {author} {\bibfnamefont {D.}~\bibnamefont
  {Dong}}, \bibinfo {author} {\bibfnamefont {C.}~\bibnamefont {Chen}}, \bibinfo
  {author} {\bibfnamefont {H.}~\bibnamefont {Li}},\ and\ \bibinfo {author}
  {\bibfnamefont {T.-J.}\ \bibnamefont {Tarn}},\ }\bibfield  {title} {\bibinfo
  {title} {Quantum reinforcement learning},\ }\href@noop {} {\bibfield
  {journal} {\bibinfo  {journal} {IEEE Transactions on Systems, Man, and
  Cybernetics, Part B (Cybernetics)}\ }\textbf {\bibinfo {volume} {38}},\
  \bibinfo {pages} {1207} (\bibinfo {year} {2008})}\BibitemShut {NoStop}%
\bibitem [{\citenamefont {Paparo}\ \emph {et~al.}(2014)\citenamefont {Paparo},
  \citenamefont {Dunjko}, \citenamefont {Makmal}, \citenamefont
  {Martin-Delgado},\ and\ \citenamefont {Briegel}}]{paparo2014quantum}%
  \BibitemOpen
  \bibfield  {author} {\bibinfo {author} {\bibfnamefont {G.~D.}\ \bibnamefont
  {Paparo}}, \bibinfo {author} {\bibfnamefont {V.}~\bibnamefont {Dunjko}},
  \bibinfo {author} {\bibfnamefont {A.}~\bibnamefont {Makmal}}, \bibinfo
  {author} {\bibfnamefont {M.~A.}\ \bibnamefont {Martin-Delgado}},\ and\
  \bibinfo {author} {\bibfnamefont {H.~J.}\ \bibnamefont {Briegel}},\
  }\bibfield  {title} {\bibinfo {title} {Quantum speedup for active learning
  agents},\ }\href@noop {} {\bibfield  {journal} {\bibinfo  {journal} {Physical
  Review X}\ }\textbf {\bibinfo {volume} {4}},\ \bibinfo {pages} {031002}
  (\bibinfo {year} {2014})}\BibitemShut {NoStop}%
\bibitem [{\citenamefont {Dunjko}\ and\ \citenamefont
  {Briegel}(2018)}]{dunjko2018machine}%
  \BibitemOpen
  \bibfield  {author} {\bibinfo {author} {\bibfnamefont {V.}~\bibnamefont
  {Dunjko}}\ and\ \bibinfo {author} {\bibfnamefont {H.~J.}\ \bibnamefont
  {Briegel}},\ }\bibfield  {title} {\bibinfo {title} {Machine learning \&
  artificial intelligence in the quantum domain: a review of recent progress},\
  }\href@noop {} {\bibfield  {journal} {\bibinfo  {journal} {Reports on
  Progress in Physics}\ }\textbf {\bibinfo {volume} {81}},\ \bibinfo {pages}
  {074001} (\bibinfo {year} {2018})}\BibitemShut {NoStop}%
\bibitem [{Note1()}]{Note1}%
  \BibitemOpen
  \bibinfo {note} {Here, we assume that agents must be causal and
  non-anticipatory, such that their decisions and internal memory can only be
  based on past information, and not as-yet-unseen future information. See
  Technical Appendix A for details.}\BibitemShut {Stop}%
\bibitem [{\citenamefont {Nielsen}\ and\ \citenamefont
  {Chuang}(2000)}]{nielsen2000quantum}%
  \BibitemOpen
  \bibfield  {author} {\bibinfo {author} {\bibfnamefont {M.~A.}\ \bibnamefont
  {Nielsen}}\ and\ \bibinfo {author} {\bibfnamefont {I.}~\bibnamefont
  {Chuang}},\ }\href@noop {} {\bibinfo {title} {Quantum {C}omputation and
  {Q}uantum {I}nformation}} (\bibinfo {year} {2000})\BibitemShut {NoStop}%
\bibitem [{\citenamefont {Crutchfield}\ and\ \citenamefont
  {Young}(1989)}]{crutchfield1989inferring}%
  \BibitemOpen
  \bibfield  {author} {\bibinfo {author} {\bibfnamefont {J.~P.}\ \bibnamefont
  {Crutchfield}}\ and\ \bibinfo {author} {\bibfnamefont {K.}~\bibnamefont
  {Young}},\ }\bibfield  {title} {\bibinfo {title} {Inferring statistical
  complexity},\ }\href@noop {} {\bibfield  {journal} {\bibinfo  {journal}
  {Physical Review Letters}\ }\textbf {\bibinfo {volume} {63}},\ \bibinfo
  {pages} {105} (\bibinfo {year} {1989})}\BibitemShut {NoStop}%
\bibitem [{\citenamefont {Shalizi}\ and\ \citenamefont
  {Crutchfield}(2001)}]{shalizi2001computational}%
  \BibitemOpen
  \bibfield  {author} {\bibinfo {author} {\bibfnamefont {C.~R.}\ \bibnamefont
  {Shalizi}}\ and\ \bibinfo {author} {\bibfnamefont {J.~P.}\ \bibnamefont
  {Crutchfield}},\ }\bibfield  {title} {\bibinfo {title} {Computational
  mechanics: {P}attern and prediction, structure and simplicity},\ }\href@noop
  {} {\bibfield  {journal} {\bibinfo  {journal} {Journal of Statistical
  Physics}\ }\textbf {\bibinfo {volume} {104}},\ \bibinfo {pages} {817}
  (\bibinfo {year} {2001})}\BibitemShut {NoStop}%
\bibitem [{\citenamefont {Crutchfield}(2012)}]{crutchfield2012between}%
  \BibitemOpen
  \bibfield  {author} {\bibinfo {author} {\bibfnamefont {J.~P.}\ \bibnamefont
  {Crutchfield}},\ }\bibfield  {title} {\bibinfo {title} {Between order and
  chaos},\ }\href@noop {} {\bibfield  {journal} {\bibinfo  {journal} {Nature
  Physics}\ }\textbf {\bibinfo {volume} {8}},\ \bibinfo {pages} {17} (\bibinfo
  {year} {2012})}\BibitemShut {NoStop}%
\bibitem [{\citenamefont {Marzen}\ and\ \citenamefont
  {Crutchfield}(2018)}]{marzen2018optimized}%
  \BibitemOpen
  \bibfield  {author} {\bibinfo {author} {\bibfnamefont {S.~E.}\ \bibnamefont
  {Marzen}}\ and\ \bibinfo {author} {\bibfnamefont {J.~P.}\ \bibnamefont
  {Crutchfield}},\ }\bibfield  {title} {\bibinfo {title} {Optimized bacteria
  are environmental prediction engines},\ }\href@noop {} {\bibfield  {journal}
  {\bibinfo  {journal} {Physical Review E}\ }\textbf {\bibinfo {volume} {98}},\
  \bibinfo {pages} {012408} (\bibinfo {year} {2018})}\BibitemShut {NoStop}%
\bibitem [{\citenamefont {Zhang}\ \emph {et~al.}(2019)\citenamefont {Zhang},
  \citenamefont {Lipton}, \citenamefont {Pineda}, \citenamefont
  {Azizzadenesheli}, \citenamefont {Anandkumar}, \citenamefont {Itti},
  \citenamefont {Pineau},\ and\ \citenamefont
  {Furlanello}}]{zhang2019learning}%
  \BibitemOpen
  \bibfield  {author} {\bibinfo {author} {\bibfnamefont {A.}~\bibnamefont
  {Zhang}}, \bibinfo {author} {\bibfnamefont {Z.~C.}\ \bibnamefont {Lipton}},
  \bibinfo {author} {\bibfnamefont {L.}~\bibnamefont {Pineda}}, \bibinfo
  {author} {\bibfnamefont {K.}~\bibnamefont {Azizzadenesheli}}, \bibinfo
  {author} {\bibfnamefont {A.}~\bibnamefont {Anandkumar}}, \bibinfo {author}
  {\bibfnamefont {L.}~\bibnamefont {Itti}}, \bibinfo {author} {\bibfnamefont
  {J.}~\bibnamefont {Pineau}},\ and\ \bibinfo {author} {\bibfnamefont
  {T.}~\bibnamefont {Furlanello}},\ }\bibfield  {title} {\bibinfo {title}
  {Learning causal state representations of partially observable
  environments},\ }\href@noop {} {\bibfield  {journal} {\bibinfo  {journal}
  {arXiv:1906.10437}\ } (\bibinfo {year} {2019})}\BibitemShut {NoStop}%
\bibitem [{\citenamefont {Boyd}\ \emph {et~al.}(2017)\citenamefont {Boyd},
  \citenamefont {Mandal},\ and\ \citenamefont
  {Crutchfield}}]{boyd2017leveraging}%
  \BibitemOpen
  \bibfield  {author} {\bibinfo {author} {\bibfnamefont {A.~B.}\ \bibnamefont
  {Boyd}}, \bibinfo {author} {\bibfnamefont {D.}~\bibnamefont {Mandal}},\ and\
  \bibinfo {author} {\bibfnamefont {J.~P.}\ \bibnamefont {Crutchfield}},\
  }\bibfield  {title} {\bibinfo {title} {Leveraging environmental correlations:
  {T}he thermodynamics of requisite variety},\ }\href@noop {} {\bibfield
  {journal} {\bibinfo  {journal} {Journal of Statistical Physics}\ }\textbf
  {\bibinfo {volume} {167}},\ \bibinfo {pages} {1555} (\bibinfo {year}
  {2017})}\BibitemShut {NoStop}%
\bibitem [{\citenamefont {Cabello}\ \emph {et~al.}(2018)\citenamefont
  {Cabello}, \citenamefont {Gu}, \citenamefont {G{\"u}hne},\ and\ \citenamefont
  {Xu}}]{cabello2018optimal}%
  \BibitemOpen
  \bibfield  {author} {\bibinfo {author} {\bibfnamefont {A.}~\bibnamefont
  {Cabello}}, \bibinfo {author} {\bibfnamefont {M.}~\bibnamefont {Gu}},
  \bibinfo {author} {\bibfnamefont {O.}~\bibnamefont {G{\"u}hne}},\ and\
  \bibinfo {author} {\bibfnamefont {Z.-P.}\ \bibnamefont {Xu}},\ }\bibfield
  {title} {\bibinfo {title} {Optimal classical simulation of state-independent
  quantum contextuality},\ }\href@noop {} {\bibfield  {journal} {\bibinfo
  {journal} {Physical Review Letters}\ }\textbf {\bibinfo {volume} {120}},\
  \bibinfo {pages} {130401} (\bibinfo {year} {2018})}\BibitemShut {NoStop}%
\bibitem [{\citenamefont {Suen}\ \emph {et~al.}(2017)\citenamefont {Suen},
  \citenamefont {Thompson}, \citenamefont {Garner}, \citenamefont {Vedral},\
  and\ \citenamefont {Gu}}]{suen2017classical}%
  \BibitemOpen
  \bibfield  {author} {\bibinfo {author} {\bibfnamefont {W.~Y.}\ \bibnamefont
  {Suen}}, \bibinfo {author} {\bibfnamefont {J.}~\bibnamefont {Thompson}},
  \bibinfo {author} {\bibfnamefont {A.~J.~P.}\ \bibnamefont {Garner}}, \bibinfo
  {author} {\bibfnamefont {V.}~\bibnamefont {Vedral}},\ and\ \bibinfo {author}
  {\bibfnamefont {M.}~\bibnamefont {Gu}},\ }\bibfield  {title} {\bibinfo
  {title} {The classical-quantum divergence of complexity in modelling spin
  chains},\ }\href@noop {} {\bibfield  {journal} {\bibinfo  {journal}
  {Quantum}\ }\textbf {\bibinfo {volume} {1}},\ \bibinfo {pages} {25} (\bibinfo
  {year} {2017})}\BibitemShut {NoStop}%
\bibitem [{\citenamefont {Dennery}\ and\ \citenamefont
  {Krzywicki}(1996)}]{dennery1996mathematics}%
  \BibitemOpen
  \bibfield  {author} {\bibinfo {author} {\bibfnamefont {P.}~\bibnamefont
  {Dennery}}\ and\ \bibinfo {author} {\bibfnamefont {A.}~\bibnamefont
  {Krzywicki}},\ }\href@noop {} {\emph {\bibinfo {title} {Mathematics for
  Physicists}}},\ Dover Books on Physics Series\ (\bibinfo  {publisher} {Dover
  Publications},\ \bibinfo {year} {1996})\BibitemShut {NoStop}%
\bibitem [{\citenamefont {Marzen}\ and\ \citenamefont
  {Crutchfield}(2015)}]{marzen2015informational}%
  \BibitemOpen
  \bibfield  {author} {\bibinfo {author} {\bibfnamefont {S.~E.}\ \bibnamefont
  {Marzen}}\ and\ \bibinfo {author} {\bibfnamefont {J.~P.}\ \bibnamefont
  {Crutchfield}},\ }\bibfield  {title} {\bibinfo {title} {Informational and
  causal architecture of discrete-time renewal processes},\ }\href@noop {}
  {\bibfield  {journal} {\bibinfo  {journal} {Entropy}\ }\textbf {\bibinfo
  {volume} {17}},\ \bibinfo {pages} {4891} (\bibinfo {year}
  {2015})}\BibitemShut {NoStop}%
\bibitem [{\citenamefont {Woods}\ \emph {et~al.}(2018)\citenamefont {Woods},
  \citenamefont {Silva}, \citenamefont {P{\"u}tz}, \citenamefont {Stupar},\
  and\ \citenamefont {Renner}}]{woods2018quantum}%
  \BibitemOpen
  \bibfield  {author} {\bibinfo {author} {\bibfnamefont {M.~P.}\ \bibnamefont
  {Woods}}, \bibinfo {author} {\bibfnamefont {R.}~\bibnamefont {Silva}},
  \bibinfo {author} {\bibfnamefont {G.}~\bibnamefont {P{\"u}tz}}, \bibinfo
  {author} {\bibfnamefont {S.}~\bibnamefont {Stupar}},\ and\ \bibinfo {author}
  {\bibfnamefont {R.}~\bibnamefont {Renner}},\ }\bibfield  {title} {\bibinfo
  {title} {Quantum clocks are more accurate than classical ones},\ }\href@noop
  {} {\bibfield  {journal} {\bibinfo  {journal} {arXiv:1806.00491}\ } (\bibinfo
  {year} {2018})}\BibitemShut {NoStop}%
\bibitem [{\citenamefont {Yang}\ and\ \citenamefont
  {Renner}(2020)}]{yang2020ultimate}%
  \BibitemOpen
  \bibfield  {author} {\bibinfo {author} {\bibfnamefont {Y.}~\bibnamefont
  {Yang}}\ and\ \bibinfo {author} {\bibfnamefont {R.}~\bibnamefont {Renner}},\
  }\bibfield  {title} {\bibinfo {title} {Ultimate limit on time signal
  generation},\ }\href@noop {} {\bibfield  {journal} {\bibinfo  {journal}
  {arXiv:2004.07857}\ } (\bibinfo {year} {2020})}\BibitemShut {NoStop}%
\bibitem [{\citenamefont {Smith}(1958)}]{smith1958renewal}%
  \BibitemOpen
  \bibfield  {author} {\bibinfo {author} {\bibfnamefont {W.~L.}\ \bibnamefont
  {Smith}},\ }\bibfield  {title} {\bibinfo {title} {Renewal theory and its
  ramifications},\ }\href@noop {} {\bibfield  {journal} {\bibinfo  {journal}
  {Journal of the Royal Statistical Society. Series B (Methodological)}\ ,\
  \bibinfo {pages} {243}} (\bibinfo {year} {1958})}\BibitemShut {NoStop}%
\bibitem [{Note2()}]{Note2}%
  \BibitemOpen
  \bibinfo {note} {Note that though we are considering the memory costs with
  respect to a particular input strategy, we drop the $\protect \mathcal {R}$
  from the subscript for notational brevity.}\BibitemShut {Stop}%
\bibitem [{\citenamefont {Kaelbling}\ \emph {et~al.}(1996)\citenamefont
  {Kaelbling}, \citenamefont {Littman},\ and\ \citenamefont
  {Moore}}]{kaelbling1996reinforcement}%
  \BibitemOpen
  \bibfield  {author} {\bibinfo {author} {\bibfnamefont {L.~P.}\ \bibnamefont
  {Kaelbling}}, \bibinfo {author} {\bibfnamefont {M.~L.}\ \bibnamefont
  {Littman}},\ and\ \bibinfo {author} {\bibfnamefont {A.~W.}\ \bibnamefont
  {Moore}},\ }\bibfield  {title} {\bibinfo {title} {Reinforcement learning: {A}
  survey},\ }\href@noop {} {\bibfield  {journal} {\bibinfo  {journal} {Journal
  of artificial intelligence research}\ }\textbf {\bibinfo {volume} {4}},\
  \bibinfo {pages} {237} (\bibinfo {year} {1996})}\BibitemShut {NoStop}%
\bibitem [{\citenamefont {Bu}\ \emph {et~al.}(2008)\citenamefont {Bu},
  \citenamefont {Babu}, \citenamefont {De~Schutter} \emph
  {et~al.}}]{bu2008comprehensive}%
  \BibitemOpen
  \bibfield  {author} {\bibinfo {author} {\bibfnamefont {L.}~\bibnamefont
  {Bu}}, \bibinfo {author} {\bibfnamefont {R.}~\bibnamefont {Babu}}, \bibinfo
  {author} {\bibfnamefont {B.}~\bibnamefont {De~Schutter}}, \emph {et~al.},\
  }\bibfield  {title} {\bibinfo {title} {A comprehensive survey of multiagent
  reinforcement learning},\ }\href@noop {} {\bibfield  {journal} {\bibinfo
  {journal} {IEEE Transactions on Systems, Man, and Cybernetics, Part C
  (Applications and Reviews)}\ }\textbf {\bibinfo {volume} {38}},\ \bibinfo
  {pages} {156} (\bibinfo {year} {2008})}\BibitemShut {NoStop}%
\bibitem [{\citenamefont {Kober}\ \emph {et~al.}(2013)\citenamefont {Kober},
  \citenamefont {Bagnell},\ and\ \citenamefont
  {Peters}}]{kober2013reinforcement}%
  \BibitemOpen
  \bibfield  {author} {\bibinfo {author} {\bibfnamefont {J.}~\bibnamefont
  {Kober}}, \bibinfo {author} {\bibfnamefont {J.~A.}\ \bibnamefont {Bagnell}},\
  and\ \bibinfo {author} {\bibfnamefont {J.}~\bibnamefont {Peters}},\
  }\bibfield  {title} {\bibinfo {title} {Reinforcement learning in robotics: A
  survey},\ }\href@noop {} {\bibfield  {journal} {\bibinfo  {journal} {The
  International Journal of Robotics Research}\ }\textbf {\bibinfo {volume}
  {32}},\ \bibinfo {pages} {1238} (\bibinfo {year} {2013})}\BibitemShut
  {NoStop}%
\bibitem [{\citenamefont {Sutton}\ and\ \citenamefont
  {Barto}(2018)}]{sutton2018reinforcement}%
  \BibitemOpen
  \bibfield  {author} {\bibinfo {author} {\bibfnamefont {R.~S.}\ \bibnamefont
  {Sutton}}\ and\ \bibinfo {author} {\bibfnamefont {A.~G.}\ \bibnamefont
  {Barto}},\ }\href@noop {} {\emph {\bibinfo {title} {Reinforcement learning:
  An introduction}}}\ (\bibinfo  {publisher} {MIT press},\ \bibinfo {year}
  {2018})\BibitemShut {NoStop}%
\bibitem [{\citenamefont {Ghafari}\ \emph
  {et~al.}(2019{\natexlab{a}})\citenamefont {Ghafari}, \citenamefont
  {Tischler}, \citenamefont {Thompson}, \citenamefont {Gu}, \citenamefont
  {Shalm}, \citenamefont {Verma}, \citenamefont {Nam}, \citenamefont {Patel},
  \citenamefont {Wiseman},\ and\ \citenamefont {Pryde}}]{ghafari2018single}%
  \BibitemOpen
  \bibfield  {author} {\bibinfo {author} {\bibfnamefont {F.}~\bibnamefont
  {Ghafari}}, \bibinfo {author} {\bibfnamefont {N.}~\bibnamefont {Tischler}},
  \bibinfo {author} {\bibfnamefont {J.}~\bibnamefont {Thompson}}, \bibinfo
  {author} {\bibfnamefont {M.}~\bibnamefont {Gu}}, \bibinfo {author}
  {\bibfnamefont {L.~K.}\ \bibnamefont {Shalm}}, \bibinfo {author}
  {\bibfnamefont {V.~B.}\ \bibnamefont {Verma}}, \bibinfo {author}
  {\bibfnamefont {S.~W.}\ \bibnamefont {Nam}}, \bibinfo {author} {\bibfnamefont
  {R.~B.}\ \bibnamefont {Patel}}, \bibinfo {author} {\bibfnamefont {H.~M.}\
  \bibnamefont {Wiseman}},\ and\ \bibinfo {author} {\bibfnamefont {G.~J.}\
  \bibnamefont {Pryde}},\ }\bibfield  {title} {\bibinfo {title} {Dimensional
  quantum memory advantage in the simulation of stochastic processes},\
  }\href@noop {} {\bibfield  {journal} {\bibinfo  {journal} {Physical Review
  X}\ }\textbf {\bibinfo {volume} {9}},\ \bibinfo {pages} {041013} (\bibinfo
  {year} {2019}{\natexlab{a}})}\BibitemShut {NoStop}%
\bibitem [{\citenamefont {Ghafari}\ \emph
  {et~al.}(2019{\natexlab{b}})\citenamefont {Ghafari}, \citenamefont
  {Tischler}, \citenamefont {Di~Franco}, \citenamefont {Thompson},
  \citenamefont {Gu},\ and\ \citenamefont {Pryde}}]{ghafari2019interfering}%
  \BibitemOpen
  \bibfield  {author} {\bibinfo {author} {\bibfnamefont {F.}~\bibnamefont
  {Ghafari}}, \bibinfo {author} {\bibfnamefont {N.}~\bibnamefont {Tischler}},
  \bibinfo {author} {\bibfnamefont {C.}~\bibnamefont {Di~Franco}}, \bibinfo
  {author} {\bibfnamefont {J.}~\bibnamefont {Thompson}}, \bibinfo {author}
  {\bibfnamefont {M.}~\bibnamefont {Gu}},\ and\ \bibinfo {author}
  {\bibfnamefont {G.~J.}\ \bibnamefont {Pryde}},\ }\bibfield  {title} {\bibinfo
  {title} {Interfering trajectories in experimental quantum-enhanced stochastic
  simulation},\ }\href@noop {} {\bibfield  {journal} {\bibinfo  {journal}
  {Nature Communications}\ }\textbf {\bibinfo {volume} {10}},\ \bibinfo {pages}
  {1630} (\bibinfo {year} {2019}{\natexlab{b}})}\BibitemShut {NoStop}%
\bibitem [{\citenamefont {Gilks}\ \emph {et~al.}(1995)\citenamefont {Gilks},
  \citenamefont {Richardson},\ and\ \citenamefont
  {Spiegelhalter}}]{gilks1995markov}%
  \BibitemOpen
  \bibfield  {author} {\bibinfo {author} {\bibfnamefont {W.~R.}\ \bibnamefont
  {Gilks}}, \bibinfo {author} {\bibfnamefont {S.}~\bibnamefont {Richardson}},\
  and\ \bibinfo {author} {\bibfnamefont {D.}~\bibnamefont {Spiegelhalter}},\
  }\href@noop {} {\emph {\bibinfo {title} {{M}arkov {C}hain {M}onte {C}arlo in
  {P}ractice}}},\ Chapman \& Hall/CRC Interdisciplinary Statistics\ (\bibinfo
  {publisher} {Taylor \& Francis},\ \bibinfo {year} {1995})\BibitemShut
  {NoStop}%
\bibitem [{\citenamefont {Loomis}\ and\ \citenamefont
  {Crutchfield}(2019)}]{loomis2019strong}%
  \BibitemOpen
  \bibfield  {author} {\bibinfo {author} {\bibfnamefont {S.~P.}\ \bibnamefont
  {Loomis}}\ and\ \bibinfo {author} {\bibfnamefont {J.~P.}\ \bibnamefont
  {Crutchfield}},\ }\bibfield  {title} {\bibinfo {title} {Strong and weak
  optimizations in classical and quantum models of stochastic processes},\
  }\href@noop {} {\bibfield  {journal} {\bibinfo  {journal} {Journal of
  Statistical Physics}\ }\textbf {\bibinfo {volume} {176}},\ \bibinfo {pages}
  {1317} (\bibinfo {year} {2019})}\BibitemShut {NoStop}%
\bibitem [{\citenamefont {Elliott}(2021{\natexlab{a}})}]{elliott2021quantum}%
  \BibitemOpen
  \bibfield  {author} {\bibinfo {author} {\bibfnamefont {T.~J.}\ \bibnamefont
  {Elliott}},\ }\bibfield  {title} {\bibinfo {title} {{Quantum coarse graining
  for extreme dimension reduction in modeling stochastic temporal dynamics}},\
  }\href@noop {} {\bibfield  {journal} {\bibinfo  {journal} {PRX Quantum}\
  }\textbf {\bibinfo {volume} {2}},\ \bibinfo {pages} {020342} (\bibinfo {year}
  {2021}{\natexlab{a}})}\BibitemShut {NoStop}%
\bibitem [{\citenamefont {Cabello}\ \emph {et~al.}(2016)\citenamefont
  {Cabello}, \citenamefont {Gu}, \citenamefont {G{\"u}hne}, \citenamefont
  {Larsson},\ and\ \citenamefont {Wiesner}}]{cabello2016thermodynamical}%
  \BibitemOpen
  \bibfield  {author} {\bibinfo {author} {\bibfnamefont {A.}~\bibnamefont
  {Cabello}}, \bibinfo {author} {\bibfnamefont {M.}~\bibnamefont {Gu}},
  \bibinfo {author} {\bibfnamefont {O.}~\bibnamefont {G{\"u}hne}}, \bibinfo
  {author} {\bibfnamefont {J.-{\AA}.}\ \bibnamefont {Larsson}},\ and\ \bibinfo
  {author} {\bibfnamefont {K.}~\bibnamefont {Wiesner}},\ }\bibfield  {title}
  {\bibinfo {title} {Thermodynamical cost of some interpretations of quantum
  theory},\ }\href@noop {} {\bibfield  {journal} {\bibinfo  {journal} {Physical
  Review A}\ }\textbf {\bibinfo {volume} {94}},\ \bibinfo {pages} {052127}
  (\bibinfo {year} {2016})}\BibitemShut {NoStop}%
\bibitem [{\citenamefont {Loomis}\ and\ \citenamefont
  {Crutchfield}(2020)}]{loomis2020thermal}%
  \BibitemOpen
  \bibfield  {author} {\bibinfo {author} {\bibfnamefont {S.~P.}\ \bibnamefont
  {Loomis}}\ and\ \bibinfo {author} {\bibfnamefont {J.~P.}\ \bibnamefont
  {Crutchfield}},\ }\bibfield  {title} {\bibinfo {title} {Thermal efficiency of
  quantum memory compression},\ }\href@noop {} {\bibfield  {journal} {\bibinfo
  {journal} {Physical Review Letters}\ }\textbf {\bibinfo {volume} {125}},\
  \bibinfo {pages} {020601} (\bibinfo {year} {2020})}\BibitemShut {NoStop}%
\bibitem [{\citenamefont {Elliott}(2021{\natexlab{b}})}]{elliott2021memory}%
  \BibitemOpen
  \bibfield  {author} {\bibinfo {author} {\bibfnamefont {T.~J.}\ \bibnamefont
  {Elliott}},\ }\bibfield  {title} {\bibinfo {title} {Memory compression and
  thermal efficiency of quantum implementations of nondeterministic hidden
  {M}arkov models},\ }\href@noop {} {\bibfield  {journal} {\bibinfo  {journal}
  {Physical Review A}\ }\textbf {\bibinfo {volume} {103}},\ \bibinfo {pages}
  {052615} (\bibinfo {year} {2021}{\natexlab{b}})}\BibitemShut {NoStop}%
\bibitem [{\citenamefont {Ho}\ \emph {et~al.}(2020)\citenamefont {Ho},
  \citenamefont {Gu},\ and\ \citenamefont {Elliott}}]{ho2020robust}%
  \BibitemOpen
  \bibfield  {author} {\bibinfo {author} {\bibfnamefont {M.}~\bibnamefont
  {Ho}}, \bibinfo {author} {\bibfnamefont {M.}~\bibnamefont {Gu}},\ and\
  \bibinfo {author} {\bibfnamefont {T.~J.}\ \bibnamefont {Elliott}},\
  }\bibfield  {title} {\bibinfo {title} {Robust inference of memory structure
  for efficient quantum modeling of stochastic processes},\ }\href@noop {}
  {\bibfield  {journal} {\bibinfo  {journal} {Physical Review A}\ }\textbf
  {\bibinfo {volume} {101}},\ \bibinfo {pages} {032327} (\bibinfo {year}
  {2020})}\BibitemShut {NoStop}%
\bibitem [{\citenamefont {Yang}\ \emph {et~al.}(2020)\citenamefont {Yang},
  \citenamefont {Binder}, \citenamefont {Gu},\ and\ \citenamefont
  {Elliott}}]{yang2020measures}%
  \BibitemOpen
  \bibfield  {author} {\bibinfo {author} {\bibfnamefont {C.}~\bibnamefont
  {Yang}}, \bibinfo {author} {\bibfnamefont {F.~C.}\ \bibnamefont {Binder}},
  \bibinfo {author} {\bibfnamefont {M.}~\bibnamefont {Gu}},\ and\ \bibinfo
  {author} {\bibfnamefont {T.~J.}\ \bibnamefont {Elliott}},\ }\bibfield
  {title} {\bibinfo {title} {Measures of distinguishability between stochastic
  processes},\ }\href@noop {} {\bibfield  {journal} {\bibinfo  {journal}
  {Physical Review E}\ }\textbf {\bibinfo {volume} {101}},\ \bibinfo {pages}
  {062137} (\bibinfo {year} {2020})}\BibitemShut {NoStop}%
\bibitem [{\citenamefont {Flitney}\ \emph {et~al.}(2004)\citenamefont
  {Flitney}, \citenamefont {Abbott},\ and\ \citenamefont
  {Johnson}}]{flitney2004quantum}%
  \BibitemOpen
  \bibfield  {author} {\bibinfo {author} {\bibfnamefont {A.~P.}\ \bibnamefont
  {Flitney}}, \bibinfo {author} {\bibfnamefont {D.}~\bibnamefont {Abbott}},\
  and\ \bibinfo {author} {\bibfnamefont {N.~F.}\ \bibnamefont {Johnson}},\
  }\bibfield  {title} {\bibinfo {title} {Quantum walks with history
  dependence},\ }\href@noop {} {\bibfield  {journal} {\bibinfo  {journal}
  {Journal of Physics A: Mathematical and General}\ }\textbf {\bibinfo {volume}
  {37}},\ \bibinfo {pages} {7581} (\bibinfo {year} {2004})}\BibitemShut
  {NoStop}%
\bibitem [{\citenamefont {McGettrick}(2009)}]{mcgettrick2009one}%
  \BibitemOpen
  \bibfield  {author} {\bibinfo {author} {\bibfnamefont {M.}~\bibnamefont
  {McGettrick}},\ }\bibfield  {title} {\bibinfo {title} {One dimensional
  quantum walks with memory},\ }\href@noop {} {\bibfield  {journal} {\bibinfo
  {journal} {arXiv:0911.1653}\ } (\bibinfo {year} {2009})}\BibitemShut
  {NoStop}%
\bibitem [{\citenamefont {Rohde}\ \emph {et~al.}(2013)\citenamefont {Rohde},
  \citenamefont {Brennen},\ and\ \citenamefont {Gilchrist}}]{rohde2013quantum}%
  \BibitemOpen
  \bibfield  {author} {\bibinfo {author} {\bibfnamefont {P.~P.}\ \bibnamefont
  {Rohde}}, \bibinfo {author} {\bibfnamefont {G.~K.}\ \bibnamefont {Brennen}},\
  and\ \bibinfo {author} {\bibfnamefont {A.}~\bibnamefont {Gilchrist}},\
  }\bibfield  {title} {\bibinfo {title} {Quantum walks with memory provided by
  recycled coins and a memory of the coin-flip history},\ }\href@noop {}
  {\bibfield  {journal} {\bibinfo  {journal} {Physical Review A}\ }\textbf
  {\bibinfo {volume} {87}},\ \bibinfo {pages} {052302} (\bibinfo {year}
  {2013})}\BibitemShut {NoStop}%
\bibitem [{\citenamefont {Rozema}\ \emph {et~al.}(2014)\citenamefont {Rozema},
  \citenamefont {Mahler}, \citenamefont {Hayat}, \citenamefont {Turner},\ and\
  \citenamefont {Steinberg}}]{rozema2014quantum}%
  \BibitemOpen
  \bibfield  {author} {\bibinfo {author} {\bibfnamefont {L.~A.}\ \bibnamefont
  {Rozema}}, \bibinfo {author} {\bibfnamefont {D.~H.}\ \bibnamefont {Mahler}},
  \bibinfo {author} {\bibfnamefont {A.}~\bibnamefont {Hayat}}, \bibinfo
  {author} {\bibfnamefont {P.~S.}\ \bibnamefont {Turner}},\ and\ \bibinfo
  {author} {\bibfnamefont {A.~M.}\ \bibnamefont {Steinberg}},\ }\bibfield
  {title} {\bibinfo {title} {Quantum data compression of a qubit ensemble},\
  }\href@noop {} {\bibfield  {journal} {\bibinfo  {journal} {Physical Review
  Letters}\ }\textbf {\bibinfo {volume} {113}},\ \bibinfo {pages} {160504}
  (\bibinfo {year} {2014})}\BibitemShut {NoStop}%
\bibitem [{\citenamefont {Yang}\ \emph {et~al.}(2018)\citenamefont {Yang},
  \citenamefont {Chiribella},\ and\ \citenamefont {Hayashi}}]{yang2018quantum}%
  \BibitemOpen
  \bibfield  {author} {\bibinfo {author} {\bibfnamefont {Y.}~\bibnamefont
  {Yang}}, \bibinfo {author} {\bibfnamefont {G.}~\bibnamefont {Chiribella}},\
  and\ \bibinfo {author} {\bibfnamefont {M.}~\bibnamefont {Hayashi}},\
  }\bibfield  {title} {\bibinfo {title} {Quantum stopwatch: how to store time
  in a quantum memory},\ }\href@noop {} {\bibfield  {journal} {\bibinfo
  {journal} {Proceedings of the Royal Society A: Mathematical, Physical and
  Engineering Sciences}\ }\textbf {\bibinfo {volume} {474}},\ \bibinfo {pages}
  {20170773} (\bibinfo {year} {2018})}\BibitemShut {NoStop}%
\bibitem [{\citenamefont {Romero}\ \emph {et~al.}(2017)\citenamefont {Romero},
  \citenamefont {Olson},\ and\ \citenamefont
  {Aspuru-Guzik}}]{romero2017quantum}%
  \BibitemOpen
  \bibfield  {author} {\bibinfo {author} {\bibfnamefont {J.}~\bibnamefont
  {Romero}}, \bibinfo {author} {\bibfnamefont {J.~P.}\ \bibnamefont {Olson}},\
  and\ \bibinfo {author} {\bibfnamefont {A.}~\bibnamefont {Aspuru-Guzik}},\
  }\bibfield  {title} {\bibinfo {title} {Quantum autoencoders for efficient
  compression of quantum data},\ }\href@noop {} {\bibfield  {journal} {\bibinfo
   {journal} {Quantum Science and Technology}\ }\textbf {\bibinfo {volume}
  {2}},\ \bibinfo {pages} {045001} (\bibinfo {year} {2017})}\BibitemShut
  {NoStop}%
\bibitem [{\citenamefont {Wan}\ \emph {et~al.}(2017)\citenamefont {Wan},
  \citenamefont {Dahlsten}, \citenamefont {Kristj{\'a}nsson}, \citenamefont
  {Gardner},\ and\ \citenamefont {Kim}}]{wan2017quantum}%
  \BibitemOpen
  \bibfield  {author} {\bibinfo {author} {\bibfnamefont {K.~H.}\ \bibnamefont
  {Wan}}, \bibinfo {author} {\bibfnamefont {O.}~\bibnamefont {Dahlsten}},
  \bibinfo {author} {\bibfnamefont {H.}~\bibnamefont {Kristj{\'a}nsson}},
  \bibinfo {author} {\bibfnamefont {R.}~\bibnamefont {Gardner}},\ and\ \bibinfo
  {author} {\bibfnamefont {M.~S.}\ \bibnamefont {Kim}},\ }\bibfield  {title}
  {\bibinfo {title} {Quantum generalisation of feedforward neural networks},\
  }\href@noop {} {\bibfield  {journal} {\bibinfo  {journal} {npj Quantum
  Information}\ }\textbf {\bibinfo {volume} {3}},\ \bibinfo {pages} {1}
  (\bibinfo {year} {2017})}\BibitemShut {NoStop}%
\bibitem [{\citenamefont {Pepper}\ \emph {et~al.}(2019)\citenamefont {Pepper},
  \citenamefont {Tischler},\ and\ \citenamefont
  {Pryde}}]{pepper2019experimental}%
  \BibitemOpen
  \bibfield  {author} {\bibinfo {author} {\bibfnamefont {A.}~\bibnamefont
  {Pepper}}, \bibinfo {author} {\bibfnamefont {N.}~\bibnamefont {Tischler}},\
  and\ \bibinfo {author} {\bibfnamefont {G.~J.}\ \bibnamefont {Pryde}},\
  }\bibfield  {title} {\bibinfo {title} {Experimental realization of a quantum
  autoencoder: The compression of qutrits via machine learning},\ }\href@noop
  {} {\bibfield  {journal} {\bibinfo  {journal} {Physical Review Letters}\
  }\textbf {\bibinfo {volume} {122}},\ \bibinfo {pages} {060501} (\bibinfo
  {year} {2019})}\BibitemShut {NoStop}%
\bibitem [{\citenamefont {Huang}\ \emph {et~al.}(2020)\citenamefont {Huang},
  \citenamefont {Ma}, \citenamefont {Yin}, \citenamefont {Tang}, \citenamefont
  {Dong}, \citenamefont {Chen}, \citenamefont {Xiang}, \citenamefont {Li},\
  and\ \citenamefont {Guo}}]{huang2019realization}%
  \BibitemOpen
  \bibfield  {author} {\bibinfo {author} {\bibfnamefont {C.-J.}\ \bibnamefont
  {Huang}}, \bibinfo {author} {\bibfnamefont {H.}~\bibnamefont {Ma}}, \bibinfo
  {author} {\bibfnamefont {Q.}~\bibnamefont {Yin}}, \bibinfo {author}
  {\bibfnamefont {J.-F.}\ \bibnamefont {Tang}}, \bibinfo {author}
  {\bibfnamefont {D.}~\bibnamefont {Dong}}, \bibinfo {author} {\bibfnamefont
  {C.}~\bibnamefont {Chen}}, \bibinfo {author} {\bibfnamefont {G.-Y.}\
  \bibnamefont {Xiang}}, \bibinfo {author} {\bibfnamefont {C.-F.}\ \bibnamefont
  {Li}},\ and\ \bibinfo {author} {\bibfnamefont {G.-C.}\ \bibnamefont {Guo}},\
  }\bibfield  {title} {\bibinfo {title} {Realization of a quantum autoencoder
  for lossless compression of quantum data},\ }\href@noop {} {\bibfield
  {journal} {\bibinfo  {journal} {Physical Review A}\ }\textbf {\bibinfo
  {volume} {102}},\ \bibinfo {pages} {032412} (\bibinfo {year}
  {2020})}\BibitemShut {NoStop}%
\bibitem [{\citenamefont {Dunjko}\ \emph {et~al.}(2016)\citenamefont {Dunjko},
  \citenamefont {Taylor},\ and\ \citenamefont {Briegel}}]{dunjko2016quantum}%
  \BibitemOpen
  \bibfield  {author} {\bibinfo {author} {\bibfnamefont {V.}~\bibnamefont
  {Dunjko}}, \bibinfo {author} {\bibfnamefont {J.~M.}\ \bibnamefont {Taylor}},\
  and\ \bibinfo {author} {\bibfnamefont {H.~J.}\ \bibnamefont {Briegel}},\
  }\bibfield  {title} {\bibinfo {title} {Quantum-enhanced machine learning},\
  }\href@noop {} {\bibfield  {journal} {\bibinfo  {journal} {Physical Review
  Letters}\ }\textbf {\bibinfo {volume} {117}},\ \bibinfo {pages} {130501}
  (\bibinfo {year} {2016})}\BibitemShut {NoStop}%
\bibitem [{\citenamefont
  {Khintchine}(1934)}]{khintchine1934korrelationstheorie}%
  \BibitemOpen
  \bibfield  {author} {\bibinfo {author} {\bibfnamefont {A.}~\bibnamefont
  {Khintchine}},\ }\bibfield  {title} {\bibinfo {title} {Korrelationstheorie
  der station{\"a}ren stochastischen {P}rozesse},\ }\href@noop {} {\bibfield
  {journal} {\bibinfo  {journal} {Mathematische Annalen}\ }\textbf {\bibinfo
  {volume} {109}},\ \bibinfo {pages} {604} (\bibinfo {year}
  {1934})}\BibitemShut {NoStop}%
\bibitem [{\citenamefont {Crutchfield}(1994)}]{crutchfield1994calculi}%
  \BibitemOpen
  \bibfield  {author} {\bibinfo {author} {\bibfnamefont {J.~P.}\ \bibnamefont
  {Crutchfield}},\ }\bibfield  {title} {\bibinfo {title} {The calculi of
  emergence: computation, dynamics and induction},\ }\href@noop {} {\bibfield
  {journal} {\bibinfo  {journal} {Physica D: Nonlinear Phenomena}\ }\textbf
  {\bibinfo {volume} {75}},\ \bibinfo {pages} {11} (\bibinfo {year}
  {1994})}\BibitemShut {NoStop}%
\bibitem [{\citenamefont {Caruso}\ \emph {et~al.}(2014)\citenamefont {Caruso},
  \citenamefont {Giovannetti}, \citenamefont {Lupo},\ and\ \citenamefont
  {Mancini}}]{caruso2014quantum}%
  \BibitemOpen
  \bibfield  {author} {\bibinfo {author} {\bibfnamefont {F.}~\bibnamefont
  {Caruso}}, \bibinfo {author} {\bibfnamefont {V.}~\bibnamefont {Giovannetti}},
  \bibinfo {author} {\bibfnamefont {C.}~\bibnamefont {Lupo}},\ and\ \bibinfo
  {author} {\bibfnamefont {S.}~\bibnamefont {Mancini}},\ }\bibfield  {title}
  {\bibinfo {title} {Quantum channels and memory effects},\ }\href@noop {}
  {\bibfield  {journal} {\bibinfo  {journal} {Reviews of Modern Physics}\
  }\textbf {\bibinfo {volume} {86}},\ \bibinfo {pages} {1203} (\bibinfo {year}
  {2014})}\BibitemShut {NoStop}%
\bibitem [{\citenamefont {Horn}\ and\ \citenamefont
  {Johnson}(1990)}]{horn1990matrix}%
  \BibitemOpen
  \bibfield  {author} {\bibinfo {author} {\bibfnamefont {R.~A.}\ \bibnamefont
  {Horn}}\ and\ \bibinfo {author} {\bibfnamefont {C.~R.}\ \bibnamefont
  {Johnson}},\ }\href@noop {} {\emph {\bibinfo {title} {Matrix Analysis}}}\
  (\bibinfo  {publisher} {Cambridge University Press},\ \bibinfo {year}
  {1990})\BibitemShut {NoStop}%
\bibitem [{\citenamefont {Neumark}(1943)}]{neumark1943spectral}%
  \BibitemOpen
  \bibfield  {author} {\bibinfo {author} {\bibfnamefont {M.}~\bibnamefont
  {Neumark}},\ }\bibfield  {title} {\bibinfo {title} {On spectral functions of
  a symmetric operator},\ }\href@noop {} {\bibfield  {journal} {\bibinfo
  {journal} {Izvestiya Rossiiskoi Akademii Nauk. Seriya Matematicheskaya}\
  }\textbf {\bibinfo {volume} {7}},\ \bibinfo {pages} {285} (\bibinfo {year}
  {1943})}\BibitemShut {NoStop}%
\bibitem [{\citenamefont {Stinespring}(1955)}]{stinespring1955positive}%
  \BibitemOpen
  \bibfield  {author} {\bibinfo {author} {\bibfnamefont {W.~F.}\ \bibnamefont
  {Stinespring}},\ }\bibfield  {title} {\bibinfo {title} {Positive functions on
  {C*}-algebras},\ }\href@noop {} {\bibfield  {journal} {\bibinfo  {journal}
  {Proceedings of the American Mathematical Society}\ }\textbf {\bibinfo
  {volume} {6}},\ \bibinfo {pages} {211} (\bibinfo {year} {1955})}\BibitemShut
  {NoStop}%
\bibitem [{\citenamefont {Akhiezer}\ and\ \citenamefont
  {Glazman}(2013)}]{akhiezer2013theory}%
  \BibitemOpen
  \bibfield  {author} {\bibinfo {author} {\bibfnamefont {N.~I.}\ \bibnamefont
  {Akhiezer}}\ and\ \bibinfo {author} {\bibfnamefont {I.~M.}\ \bibnamefont
  {Glazman}},\ }\href@noop {} {\emph {\bibinfo {title} {Theory of linear
  operators in Hilbert space}}}\ (\bibinfo  {publisher} {Courier Corporation},\
  \bibinfo {year} {2013})\BibitemShut {NoStop}%
\bibitem [{\citenamefont {Watrous}(2018)}]{watrous2018theory}%
  \BibitemOpen
  \bibfield  {author} {\bibinfo {author} {\bibfnamefont {J.}~\bibnamefont
  {Watrous}},\ }\href@noop {} {\emph {\bibinfo {title} {The Theory of Quantum
  Information}}}\ (\bibinfo  {publisher} {Cambridge University Press},\
  \bibinfo {year} {2018})\BibitemShut {NoStop}%
\bibitem [{\citenamefont {Marzen}\ and\ \citenamefont
  {Crutchfield}(2017)}]{marzen2017informational}%
  \BibitemOpen
  \bibfield  {author} {\bibinfo {author} {\bibfnamefont {S.}~\bibnamefont
  {Marzen}}\ and\ \bibinfo {author} {\bibfnamefont {J.~P.}\ \bibnamefont
  {Crutchfield}},\ }\bibfield  {title} {\bibinfo {title} {Informational and
  causal architecture of continuous-time renewal processes},\ }\href@noop {}
  {\bibfield  {journal} {\bibinfo  {journal} {Journal of Statistical Physics}\
  }\textbf {\bibinfo {volume} {168}},\ \bibinfo {pages} {109} (\bibinfo {year}
  {2017})}\BibitemShut {NoStop}%
\end{thebibliography}%

\end{document}